\documentclass[11pt, a4paper]{article}

\usepackage{lmodern}
\usepackage{amsmath,amsfonts,amssymb,amsthm}
\usepackage{mathrsfs}
\usepackage{graphics}
\usepackage{cases}
\usepackage{latexsym} % symbols
\usepackage[english]{babel}
\usepackage[utf8]{inputenc}
\usepackage[T1]{fontenc}
\usepackage{mathtools}
\usepackage{enumerate}
\usepackage{commath}
\usepackage{xcolor}

\usepackage{braket}
\usepackage{makecell}
\usepackage{comment}
\usepackage{epigraph}
\usepackage{upgreek}
\usepackage{authblk}
\usepackage{dsfont}
\usepackage{ulem}
\usepackage{stmaryrd}
\usepackage{fullpage}
\usepackage{cancel}
\usepackage{cite}
\usepackage[a4paper,vmargin={3cm,3cm},hmargin={3cm,3cm}]{geometry}
\usepackage{etoolbox}
\usepackage{thmtools}
\usepackage[bottom]{footmisc}
\usepackage{enumitem}
\usepackage[pdfpagemode=UseNone,
bookmarksopen=false,
colorlinks=true,
urlcolor=blue,
citecolor=magenta,
citebordercolor=blue,
linkcolor=blue, 
urlcolor=cyan, 
filecolor=red]{hyperref}
\usepackage{tikz}

\usepackage{titlesec}
\titleformat{\section}{\large\bf}{\thesection}{1em}{}
\titleformat{\subsection}{\bf}{\thesubsection}{1em}{}[]
\titleformat{\subsubsection}[runin]{\bf}{\thesubsubsection}{1em}{}[.]

\theoremstyle{definition}
\newtheorem{remark}{Remark}[section]

\newcommand{\R}{\mathbb{R}}

\newcommand{\I}{\ensuremath{\mathbf{i}}}

\newcommand{\la}{\lambda}

\def\note#1{\textup{\textsf{\color{blue}(#1)}}}

\renewcommand{\rho}{\varrho}
\newcommand{\Arescaled}{a}
\newcommand{\Brescaled}{b}
\newcommand{\Crescaled}{c}
\newcommand{\invgamma}{W}
\newcommand{\ratioGamma}{\Omega}
\newcommand{\ratioGammaLargeTime}{\omega}

\newcommand{\Var}{\mathrm{Var}}

\newcommand{\EE}{\ensuremath{\mathbb{E}}}

\newcommand{\be}{\begin{equation}}
\newcommand{\ee}{\end{equation}}
\newcommand{\beq}{\begin{equation}}
\newcommand{\eeq}{\end{equation}}

\def \ba {\begin{aligned}}
	\def \ea {\end{aligned}}

\newcommand{\bea}{\begin{eqnarray}}
\newcommand{\eea}{\end{eqnarray}}
\newcommand{\beqa}{\begin{eqnarray}}
\newcommand{\eeqa}{\end{eqnarray}}
\newcommand{\beqn}{\begin{eqnarray}}
\newcommand{\eeqn}{\end{eqnarray}}

\newcommand{\bc}{\begin{center}}
	\newcommand{\ec}{\end{center}}

\newcommand{\beqs}{\begin{eqnarray*}}
	\newcommand{\eeqs}{\end{eqnarray*}}

\newcommand{\bi}{\begin{itemize}}
	\newcommand{\ei}{\end{itemize}}

\DeclareMathOperator*{\argmax}{arg\,max}

\newcommand{\rmd}{\mathrm{d}}

\def\1{\mathds{1}}

\newcommand{\Det}{\mathrm{Det}}
\renewcommand{\det}{\mathrm{Det}}
\def \Pf {{\rm Pf}}

\newcommand{\Ai}{\mathrm{Ai}}

\def\<{\langle}
\def\>{\rangle}

\renewcommand*{\geq}{\geqslant}
\renewcommand*{\leq}{\leqslant}

\author[1]{\large Guillaume Barraquand}
\author[2]{\large Alexandre Krajenbrink}
\author[1]{\large Pierre Le Doussal}
\affil[1]{\small Laboratoire de Physique de l'\'Ecole Normale Sup\'erieure, ENS, Universit\'e PSL,  CNRS, Sorbonne Universit\'e, Universit\'e Paris Cité, 75005 Paris, France}
\affil[2]{\small Quantinuum and Cambridge Quantum Computing, Cambridge, United Kingdom}

\title{Half-space stationary Kardar-Parisi-Zhang equation beyond the Brownian case}
\date{}

\begin{document}
\maketitle
\begin{abstract}
    We study the Kardar-Parisi-Zhang equation on the half-line $x \geqslant 0$ with Neumann type boundary condition. 
    Stationary measures of the KPZ dynamics were characterized in recent work: they depend on two parameters, the boundary parameter $u$ of the dynamics, and the drift $-v$ of the initial condition at infinity. We consider the fluctuations of the height field when the initial condition is given by one of these stationary processes. At large time $t$, it is natural to rescale parameters as $(u,v)=t^{-1/3}(a,b)$ to study the critical region. In the special case $a+b=0$, treated in 
    previous works, the stationary process is simply Brownian. However, these Brownian stationary measures are particularly relevant in the bound phase ($a<0$) but not in the unbound phase. For instance, starting from the flat or droplet initial condition, the height field near the boundary converges to the stationary process with $a>0$ and $b=0$, which is not Brownian.  
    For $a+b\geqslant 0$, we determine exactly the large time distribution $F_{a,b}^{\rm stat}$ of the height function $h(0,t)$. As an application, 
    we obtain the exact covariance of the height field in a half-line at two times $1\ll t_1\ll t_2$ starting from stationary initial condition, as well as  estimates,  when starting from droplet initial condition, in the limit $t_1/t_2\to 1$.  
\end{abstract}

{\hypersetup{linkcolor=black}
\setcounter{tocdepth}{2}
\tableofcontents
}

\section{Introduction and main results}
\subsection{Introduction}
The Kardar-Parisi-Zhang (KPZ) equation \cite{KPZ} in one dimension describes the time evolution of the height field $h(x,t)$ 
of an interface which undergoes a local growth process driven by white noise. It is a paradigmatic element of a large universality class of
one dimensional models with identical universal behavior at large scale, the so-called KPZ class. Since the interface
is growing, with $h(x,t) \sim v_\infty t$ at large time, it is intrisically an out of equilibrium problem. One can nevertheless
ask whether a stationary state can be reached at large time. While the height at one point grows linearly in time with non trivial
$t^{1/3}$ fluctuations, the height difference between any two points, $h(x,t)-h(y,t)$, reaches a stationary distribution. Even when this distribution is known, it says nothing about the increments of the height field in time, say $h(0,t)-h(0,0)$. In this paper, we are interested in computing these temporal increments and their asymptotics  at stationarity (i.e. starting from a stationary initial condition), and apply it to compute the two-time covariance of the height field.

Let us first briefly review what is known about stationary distributions for the KPZ equation.  They depend on whether one considers the equation on the full-line, or in a restricted geometry such as a half-line or
an interval. On the full-line, it has been predicted for a long time \cite{forster1977large,parisi1990replica} 
that the KPZ equation admits the Brownian motion (BM) with an arbitrary drift as a stationary measure. This was proved rigorously in \cite{bertini1997stochastic}, 
and in \cite{hairer2018strong} for periodic boundary conditions. Interestingly, in the cases of the half-line and the interval, the generic situation is more complicated (not translation invariant, not Gaussian, see below). 
One typically imposes Neumann type boundary conditions (that is, one fixes the derivative of the height field at the boundary) so that stationary measures depend on boundary parameters. For the interval it depends on the two boundary parameters, while for the half-line it depends on one boundary parameter and on the drift at infinity. In the special case when boundary and drift parameters are such that the slope imposed at the origin has the same value as the drift at infinity, the BM (with the same drift) is again stationary, as was shown in the case of the half-line in \cite{barraquand2020half} (this specific half-line stationary measure was studied
in the equivalent directed polymer context in \cite{barraquand2021kardar}). But this Brownian stationary measure is not unique, and for arbitrary values of the  drift parameter, the stationary measures
have been found only recently.

For the KPZ equation on an interval $[0,L]$, an explicit formula for the Laplace transform (LT) of the stationary height distribution was obtained in 
\cite{corwin2021stationary} (for $L=1$, and for some range of parameters). Explicit Laplace inversion was performed shortly after
in \cite{bryc2021markov} and \cite{barraquand2021steady} (see also \cite{bryc2021markov2}). In \cite{barraquand2021steady} a rather simple and explicit
characterization of the process $h(x,t)-h(x,0)$ was obtained by two of the present authors. This characterization allowed in particular to predict the stationary measures in the limit of the half-line, letting the size of the interval to infinity. This large interval limit is actually quite non-trivial, but surprisingly, the limit of stationary measures on $[0,L]$ had been already studied in the mathematics literature \cite{hariya2004limiting}, with completely different motivations. This is why we refer below to the resulting processes on the half-line (i.e. the non-Brownian stationary distribution for KPZ in a half-space) as Hariya-Yor processes.

Let us now describe in more details the question we address in this paper. While the steady-state distribution of $h(x,t)-h(0,t)$ for the KPZ equation on the half-line $x \geq 0$ is now clear, it remains to understand the global height, that is the height at one point, for instance $h(0,t)$. Apart from situations where the boundary is very attractive, already studied in \cite{barraquand2021kardar}, the height grows at large time as $h(0,t) \simeq v_\infty t + t^{1/3} \chi$, where $\chi$ is a random variable whose distribution depends on some details of the initial condition $h(x,0)$. 
For the KPZ equation on the full-line with a stationary, i.e. BM initial condition, $\chi$ follows the Baik-Rains  distribution \cite{baik2000limiting, imamura2013stationary, imamura2012exact, borodin2015height}.  It is universal over the KPZ class and, remarkably, was measured in recent experiments on liquid crystals \cite{iwatsuka2020direct}. For the half-line, the corresponding question was addressed 
only recently, but until now only for the special case mentioned above where the BM (with drift) is still the stationary measure. The analog of the Baik-Rains 
distribution then depends on one boundary parameter, we denote it $F_a^{\rm Brownian}$, and it was obtained in \cite{betea2020stationary} (in the context of last-passage-percolation, which is equivalent by universality) and in \cite{barraquand2020half} directly for the KPZ equation
for $a=0$.

The aim of this paper is to obtain the analog of the Baik-Rains distribution for the KPZ equation on the half-line starting from
a stationary initial condition in the generic case, that is for the Hariya-Yor initial conditions. 
We show in this paper that these Hariya-Yor processes are integrable, in the sense that  one can write down simple exact formulas for the mixed exponential moments, using the framework of half-space Macdonald processes \cite{barraquand2020halfMacdonald}. Through the usual Hopf-Cole mapping $h(x,t)=\log Z(x,t)$, the process $Z(x,t)$ solves the multiplicative noise stochastic heat equation (that is the partition function of a continuous Brownian directed polymer in a random potential) and we compute explicitly all moments of $Z(x,t)$ via a Bethe ansatz type approach. %argument. 
Following a similar line as in \cite{barraquand2020half} (and previous works including \cite{borodin2016directed, AlexLD}), we express the Laplace transform of $Z(x,t)$ as Fredholm Pfaffians and determinants, which we analyze asymptotically to obtain the limiting distribution of $\chi$.

It is important to note that although our results are obtained from the large-scale analysis of the KPZ equation, we expect that they hold universally for all half-space models in the KPZ universality class. 
As an application, following a method introduced in \cite{ferrari2006scaling} for full-space models, we have obtained the two-time covariance of the KPZ height in a half-line geometry. 
These results can also be  translated in terms of the free energy of a directed polymer in the presence of a wall. 

\subsubsection*{Outline} In the following sections, we first review the stationary measures for the KPZ equation
on the half-line (Section \ref{sec:stationaryKPZ}) and discuss the various distributions that arise for the large time fluctuations of the height field (Section \ref{sec:limitingdistributionsintro}). Our main new results are presented in Section \ref{sec:introHYcase}.  We then  present an important application of the main results to the computation of two-time covariance in half-space KPZ growth in Section \ref{sec:twotimeintro}. 

The remaining sections are devoted to details of the derivations. In Section \ref{sec:moments}, we obtain the moments and Laplace transform formulas characterizing the distribution of $h(0,t)$. We analyze the formulas asymptotically in Section \ref{sec:largetime} and obtain explicit formulas for the cumulative distribution function (CDF) of the limiting distributions. We provide the details of computations of two-time covariance in half-space KPZ growth in Section \ref{sec:twotime}. Finally, in Appendix \ref{sec:Brownian}, we extend the results of our previous work \cite{barraquand2020half} to provide an explicit formula for the CDF $F_a^{\rm Brownian}$ when the boundary parameter $a\neq 0$. In Appendix  \ref{sec:xpositif}, we explain how to compute the distribution of $h(x,t)$ instead of $h(0,t)$, when $x$ is scaled of order $t^{2/3}$, and the initial condition is a Hariya-Yor processes.

\subsection{Half-space KPZ stationary measures} 
\label{sec:stationaryKPZ}
Let $Z(x,t)$ denote the solution to the half-space stochastic heat equation
\begin{equation}
\partial_t Z(x,t)= \partial_{xx} Z(x,t)+ \sqrt{2}\eta(x,t)Z(x,t),\;\;\; (x\geqslant 0, t\geqslant 0),  
\end{equation}
 with standard white noise $\eta$, initial condition $Z_0(x)$ and boundary parameter $A\in \mathbb R$, corresponding formally to $$\partial_x Z(x,t)\big\vert_{x=0}= A Z(0,t).$$ We will often denote the boundary parameter rather by the letter $u$ where $u=\frac{1}{2}+A$.  
 Then, we will say that $h_u(x,t)=\log Z(x,t)$ solves the Kardar-Parisi-Zhang equation on $\mathbb R_+$ 
 \begin{equation}
     \partial_t h_u=\partial_{xx}h_u+(\partial_x h_u)^2 +\sqrt{2} \eta
 \end{equation}
 with boundary parameter $u$.

It was noticed in \cite{barraquand2020half} that if the initial condition $x\mapsto \mapsto h_u(x,0)$ is distributed as
\begin{equation}
h_u(x,0) = B(x)+ux
\end{equation}
that is  a Brownian motion with drift $u$, then for all $t\geqslant 0$, the process 
$x\mapsto h_u(x,t)-h_u(x,0)$ has the same distribution, that is 
\begin{equation}
h_u(x,t)-h_u(0,t) \overset{(d)}{=}B(x)+ux.
\end{equation}
 In other terms, the Brownian motion with drift $u$ is a stationary distribution for the KPZ equation in a half-space with boundary parameter $u$. In infinite volume (that is for dynamics on functions of $\mathbb R$ or $\mathbb R_+$), there is no reason to expect that the stationary process is unique. Indeed, there exist other, more complicated, stationary measures for the half-space KPZ equation, recently described in \cite{barraquand2021steady}, based on results for the stationary measure on an interval from \cite{corwin2021stationary} (see also \cite{bryc2021markov, bryc2021markov2} for an equivalent description of stationary measures on an interval). These additional stationary measures depend on a parameter $v$, where $-v$ is the drift of the process at infinity, and they arise only when $u\geqslant v, v\leqslant 0$.  It is convenient to represent them on the diagram of Fig. \ref{fig:half-space}, which explains which stationary processes arise in the large time limit, depending on the boundary parameter $u$ and the drift of the initial condition. They are defined in terms of a process that we call the Hariya-Yor process, defined below and denoted $\mathcal{HY}_{u,v}$, 
\usetikzlibrary{patterns}
\usetikzlibrary{patterns.meta}
  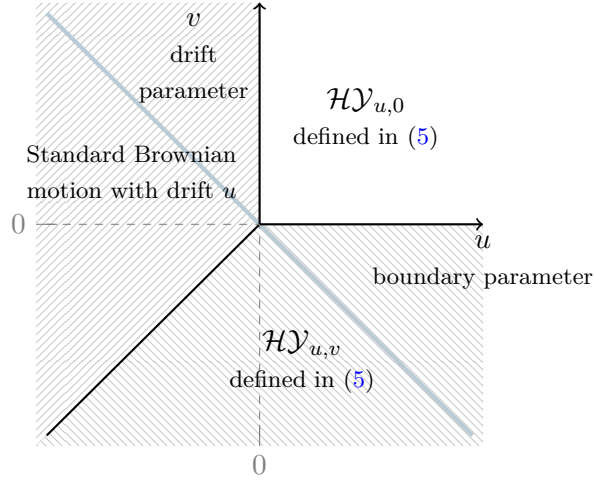
\begin{figure}
    \centering
    \begin{tikzpicture}[scale=1.4, every text node part/.style={align=center}]
    \fill[pattern=north east lines, pattern color=black!20] (-.1,-.1) --(2,2) -- (2,4.1) -- (-.1,4.1) -- cycle;
    \fill[pattern=north west lines, pattern color=black!20] (-.1,-.1) --(2,2) -- (4.1,2) -- (4.1,-.1) -- cycle;
    \draw[ultra thick, cyan!40!black!30] (0,4) -- (4,0);
    \draw[thick] (0,0) -- (2,2);
    \draw[thick, ->] (2,2) -- (2,4.1) node[anchor=north east] {$v$ \\\footnotesize   drift \\ \footnotesize  parameter};
    \draw[thick, ->] (2,2) -- (4.1,2) node[anchor=north] {$u$ \\ \footnotesize  boundary parameter};
    \draw[dashed, gray] (2,0) -- (2,2);
    \draw[dashed, gray] (0,2) -- (2,2);
    \draw[gray] (2,0.1) -- (2,-0.1) node[anchor=north] {$0$};
    \draw[gray] (0.1,2) -- (-0.1,2) node[anchor=east] {$0$};
    \draw (0.8,2.5) node{\footnotesize Standard Brownian \\\footnotesize motion with drift $u$};
    \draw (2.4,0.7) node{$\mathcal{HY}_{u,v}$ \\ \footnotesize defined in \eqref{eq:defHariyaYor}};
    \draw (3,3) node{$\mathcal{HY}_{u,0}$ \\ \footnotesize defined in \eqref{eq:defHariyaYor}};
    \end{tikzpicture}
    \caption{Phase diagram of stationary measures for the KPZ equation in the half-space $\mathbb R_+$ with boundary parameter $u$. The diagram means that if the initial condition $h(x,0)$ has drift $-v$ at infinity, the height field should converge at large time  under mild assumptions to the stationary measure indicated in one of the three regions of the $(u,v)$ plane, namely $R_1 = \{u \geq 0, v \geq 0\}$, 
    $R_2 = \{u \leq 0, u \leq v\}$ and
    $R_3 = \{u \geq v, v \leq 0\}$.
    Along the line $u+v=0$, the stationary measure is a Brownian motion with drift $u=-v$.} 
    \label{fig:half-space}
\end{figure}
 For $u> v , v\leqslant 0$ with $u+v>0$, we define the Hariya-Yor process, denoted\footnote{our notations are different from \cite{barraquand2021steady}, where a slightly different process was denoted $\mathcal{HY}_{u}^{-v}(x)$)} $\mathcal{HY}_{u,v}(x)$, by 
 \begin{equation}
   \exp\left( \mathcal{HY}_{u,v}(x) \right) := w_1  \int_{0}^x\mathrm d t e^{B_1(t)+B_2(x)-B_2(t)} + w_1w_2 e^{B_2(x)},
     \label{eq:defHariyaYor}
 \end{equation}
 where $B_1, B_2$ are independent standard Brownian motions with drifts $-v$ and $v$ respectively, and $w_1, w_2$ are independent inverse gamma random variables $w_1\sim\mathrm{Gamma}^{-1}(u+v)$ and $w_2\sim\mathrm{Gamma}^{-1}(u-v)$. We say that a random variable $X$ follows the $\mathrm{Gamma}^{-1}(\theta)$ distribution if $1/X$ is a Gamma random variable with scale parameter $1$ and shape parameter $\theta$. In other terms, $X$ is a positive random variable with density $\frac{1}{\Gamma(\theta)}\mathds{1}_{x>0} x^{-\theta}e^{-1/x}\frac{dx}{x}$.

In order to give a unified description of the stationary measures, is is convenient to define, for any $u,v\in \mathbb R$, a stationary initial condition $h_{u,v}^{\rm stat}$, where 
\begin{equation}
\label{eq:defhstat}
    h_{u,v}^{\rm stat}(x) = \begin{cases} \mathcal{HY}_{u,0}(x)- \mathcal{HY}_{u,0}(0), &\mbox{ for }u> 0, v\geqslant 0, \mbox{that is }(u,v)\in  R_1,\\
    B(x)+ ux &\mbox{ for }u\leqslant 0, v\geqslant u, \mbox{that is }(u,v)\in  R_2,\\
    \mathcal{HY}_{u,v}(x)- \mathcal{HY}_{u,v}(0), &\mbox{ for }u> v, v\leqslant 0, \mbox{that is }(u,v)\in  R_3.
    \end{cases}
\end{equation}

Let us stress that while in the phase $u> v, v\leqslant 0$ the process $h_{u,v}^{\rm stat}(x)$ describes exactly the spatial increments of $\mathcal{HY}_{u,v}(x)$, it is defined differently in the other phases and this is why we needed a new notation. We will use alternatively $h_{u,v}^{\rm stat}(x)$ and $\mathcal{HY}_{u,v}(x)$ for the following reason: $h_{u,v}^{\rm stat}(x)$ is the natural way to describe the stationary height field, in particular it is normalized so that $h_{u,v}^{\rm stat}(0)=0$. The process $\mathcal{HY}_{u,v}(x)$, however, is not normalized but contains the appropriate random shift that makes exact computations possible.

\begin{remark}
The Hariya-Yor process \eqref{eq:defHariyaYor}  is defined only when $u> v , v\leqslant 0$ with $u+v>0$. It depends on a random variable $w_1\sim \mathrm{Gamma}^{-1}(u+v)$, which explains the condition $u+v>0$. However, 
$$ h_{u,v}^{\rm stat} (x)= \mathcal{HY}_{u,v}(x)-  \mathcal{HY}_{u,v}(0) =  \mathcal{HY}_{u,v}(x) -\log w_1-\log w_2$$
depends only on $w_2\sim \mathrm{Gamma}^{-1}(u-v)$, and on two Brownian motions $B_1$, $B_2$, hence it is well-defined for any $u>v, v\leqslant 0$, without imposing the condition $u+v>0$. 
\end{remark}
\begin{remark}\label{rem:specialcaseBrownian}
For any $u> 0$, 
\begin{equation}
    \mathcal{HY}_{u,-u}(x)-\mathcal{HY}_{u,-u}(0)=B(x)+ux
\end{equation}
where $B(x)$ is a standard Brownian motion. This is a nontrivial result, discovered by Hariya and Yor in \cite{hariya2004limiting}. 
\end{remark}
\begin{remark}
As $v\to u$, it is easy to check that 
\begin{equation}
    \mathcal{HY}_{u,v}(x)-\mathcal{HY}_{u,v}(0)\xRightarrow[v\to u]{}B(x)+ux. 
\end{equation}
\end{remark}
 \begin{remark}
 Our definition of $\mathcal{HY}_{u,v}(x)$ may appear different from the process defined in \cite[Eq.~(35)]{barraquand2021steady} but it is equal up to a shift by $\log(w_1w_2)$. Indeed, the process is defined in \cite{barraquand2021steady} as 
 \begin{equation} 
H(x) = B^{(1)}(x)  + B^{(2)}(x)  \\ +\log\left( 1+\gamma_{u-v} \int_0^x e^{-2B^{(2)}(z)}dz\right),
\label{eq:halfspacehighdensity}
\end{equation}
where $B^{(1)}(x) ,  B^{(2)}(x)$ are independent Brownian motions with variance $1/2$ and drifts $0$ and $v$ respectively. Using the change of variables
$$ B^{(1)} = \frac{B_1+B_2}{2}, \;\; B^{(2)} = \frac{B_2-B_1}{2},$$
$B_1,B_2$ are two independent standard Brownian motions of drifts $-v$ and $v$ respectively,
and we have that, if we identify $\gamma_{u-v}$  in \eqref{eq:halfspacehighdensity} with $1/w_2$ in \eqref{eq:defHariyaYor}, then 
\begin{equation}
e^{H(x)} = \frac{e^{\mathcal{HY}_{u,v}(x) }}{w_1w_2},
\end{equation}
so that 
\begin{equation}
    H(x) = \mathcal{HY}_{u,v}(x)-\log(w_1) -\log( w_2). 
\end{equation}
 \end{remark}
 \begin{remark} 
 The integral \eqref{eq:defHariyaYor} is similar to the partition function of the semi-discrete O'Connell-Yor polymer \cite{o2001brownian, o2002representation} with two rows. Other O'Connell-Yor partition functions decorated by inverse Gamma variables appeared in the literature, \cite{borodin2015height, talyigas2020borodin, barraquand2021kardar}, though the partition function in \eqref{eq:defHariyaYor} is different. In the limit $u\to\infty$, we recover exactly the O'Connell-Yor partition function. 
\end{remark}

\subsection{Limiting distribution for the KPZ height on a half-line}
\label{sec:limitingdistributionsintro}
\subsubsection{Droplet initial condition}

For the droplet initial condition, i.e. $e^{h_u(x,t)} \to_{t \to 0} \delta(x)$, and boundary parameter $u$,  a phase transition occurs based on the sign of $u$. We have 
\begin{equation} \label{eq:Fdroplet} 
    \lim_{t\to \infty} \mathbb P\left( \frac{h_u(0,t)+\frac{t}{12} }{ t^{1/3} } \leqslant s \right)  = \begin{cases} 
    F_{\rm GSE}(s) &\mbox{ for }u>0,\\
    F_{\rm GOE}(s)&\mbox{ for }u=0,\\ 
   0  &\mbox{ for }u<0.
   \end{cases}
\end{equation}
In the phase $u<0$, the scaling and statistics are different: $h_u(0,t)\simeq t\left(\frac{-1}{12}+u^2\right)$ \cite{kardar1985depinning} and statistics are Gaussian on the $t^{1/2}$ scale (see  \cite{deNardisPLDTT, barraquand2021kardar}). 
If we scale $u$ close to the critical point as $u=at^{-1/3}$, we have  that 
\begin{equation} \label{eq:defFadroplet} 
    \lim_{t\to \infty} \mathbb P\left( \frac{h_{at^{-1/3}}(0,t)+\frac{t}{12} }{ t^{1/3} } \leqslant s \right)  = F^{\rm droplet}_a(s).
\end{equation}
The existence of a transition was anticipated in \cite{kardar1985depinning}. In the equivalent directed polymer problem, it corresponds to a transition to polymers bound to the wall at $x=0$ when $u<0$ to unbound polymers when $u\geqslant 0$. The statistics occurring around the phase transition \eqref{eq:Fdroplet} and the exact formula for $F^{\rm droplet}_a$ were first discovered in \cite{baik2001asymptotics, baik2001symmetrized} in the context of asymptotic fluctuations of symmetrized last passage percolation models. For the KPZ equation, these asymptotics were obtained in \cite{gueudre2012directed}  for $u=+\infty$, in \cite{borodin2016directed} for $u=1/2$, in \cite{barraquand2018stochastic} for $a=0$ (i.e. the critical case, $u=0$), and in \cite{deNardisPLDTT, AlexLD} in the general case. We have in particular $F^{\rm droplet}_0=F_{\rm GOE}$. 
\medskip 

\subsubsection{Brownian initial condition}
\label{sec:introBrowniancase}
For a Brownian initial condition $h_u(x,0)=B(x)+ux$, where $B(x)$ is a standard Brownian  motion, which is stationary for any $u\in \mathbb R$, we have 
for $u = a t^{-1/3}$
\begin{equation} \label{eq:defFaBrownian} 
\lim_{t\to \infty} \mathbb P\left( \frac{h_{at^{-1/3}}(0,t)+\frac{t}{12} }{ t^{1/3} } \leqslant s \right)  = F^{\rm Brownian}_a(s). 
\end{equation}
This statement and an exact formula for $F^{\rm Brownian}_a$  was obtained  in \cite{barraquand2020half} in the special case $a=0$. 
In the context of last-passage percolation, the analogous statement for any $a\in \mathbb R$ was shown earlier in \cite{betea2020stationary}. We also provide  in Appendix  \ref{sec:Brownian} an alternative formula for $F^{\rm Brownian}_a$, based on an extension of our earlier result for $a=0$  in \cite{barraquand2020half} (The formula for $F^{\rm Brownian}_a(s)$ appears in \eqref{eq:formulaforFaBrownian}). It remains to be shown  that  the exact formulas from \cite{betea2020half} and our formulas in \cite{barraquand2020half} and Section \ref{sec:Brownian} are equivalent. 
Note that in \cite{betea2020stationary,betea2020half}, formulas were also obtained for the height
distribution and multipoint correlations at points away from the boundary.

\medskip 
According to the phase diagram of Fig. \ref{fig:half-space}, in the bound phase, i.e. when $u<0$, for any initial condition with drift at infinity not exceeding $-u$, the spatial process $h_u(x,t)-h_u(0,t)$ should converge, as $t$ goes to infinity, to a Brownian with drift $u$, hence the importance of this case. Consequences about the geometry of the polymer and the distribution of the endpoint were investigated in \cite{barraquand2021kardar}. 

In the unbound phase $u\geqslant 0$ however, the stationary process obtained at large time is the Hariya-Yor process $\mathcal{HY}_{u,v}$, when the drift of the initial condition is positive and equals $-v>0$, and $\mathcal{HY}_{u,0}$ when the drift is negative. 
Brownian stationary measures arise only when $u+v=0$, see Remark \ref{rem:specialcaseBrownian}, which is a very special case. For droplet or flat initial condition for instance, the stationary process observed at large times is the Hariya-Yor process $\mathcal{HY}_{u,0}$. This makes the study of fluctuations starting from the Hariya-Yor initial condition particularly important in the unbound phase.  

\subsubsection{Main new result: Hariya-Yor initial condition} 
\label{sec:introHYcase}
We now focus on the regions $R_1$ and $R_3$.   We assume that the initial condition  $h_u(x,t) = \log Z(x,t)$ is given by
$h_u(x,0) = \mathcal{HY}_{u,v}(x)$, for $u> v, v\leqslant 0$ as defined in \eqref{eq:defHariyaYor}, under the additional  technical assumption $u+v>0$. We will show that, for $u=at^{-1/3}, v=bt^{-1/3}$, with $a+b>0, b\leq 0$
\begin{equation}
    \lim_{t\to \infty} \mathbb P\left( \frac{h_{at^{-1/3}}(0,t)+\frac{t}{12} }{ t^{1/3} } \leqslant s \right)  = G_{a,b}^{\rm HY}(s)
\end{equation}
where the CDF $G_{a,b}^{\rm HY}(s)$ will be explicitly determined.  An explicit formula for $G_{a,b}^{\rm HY}(s)$ is given in \eqref{eq:Fab}. When $a>0$ and $b=0$, that is in the maximal current phase, the formula simplifies and we obtain 
\begin{equation}
G_{a,0}^{\rm HY}(s) =\partial_s \left( \det(I+\tilde A_s)\left( -\frac{2}{a}+s +2R^+ \right)\right), 
\end{equation}
 where $\tilde A_s$ is an integral operator acting on $\mathbb L^2(0, +\infty)$ with  kernel 
$ \tilde A_s(x,y) =\tilde A(s+x+y)$ where
\begin{equation}
\tilde A(x)  =  \int \frac{\mathrm d z}{2\I\pi} \frac{\Arescaled +z}{\Arescaled -z}   e^{-xz+ \frac{z^3}{3}},
\label{eq:defAtildeintro}
\end{equation}
where the contour is a vertical line with real part between $0$ and $a$, and  $R^+$ is a scalar product defined using quantum mechanical notations as (see Section \ref{sec:stationarylimit} for details)
\begin{equation}
 R^{+}= \bra{ 1} \frac{\tilde A_s}{1+ \tilde A_s} \ket{ 1} . \end{equation}
%  \note{Should we include plots of this distribution function? or at least compute the variance ?! }

  \begin{remark} 
  \label{rem:atoinfty}
As $a\to+\infty$, we notice that  $G_{a,0}^{\rm HY}(s)\to F^{\rm Brownian}_{0}$ given in \cite[Eq. (7.24)]{barraquand2020half}. This limit is obvious from the formula but the reason is   non-trivial,  it can be seen as a consequence of a surprising identity in distribution obtained in \cite[Sections 4.5 and 4.6]{barraquand2020half}, see more details in Remark \ref{rem:idenityindistribution}. This identity in distribution itself comes from more general identities in distribution for the log-gamma polymer and half-space Macdonald processes found in \cite[Prop. 2.6 and Prop. 8.1]{barraquand2020halfMacdonald}.   
\end{remark}

The distribution given by $G_{a,b}^{\rm HY}$ is not centered, unlike the Baik-Rains distribution or the distribution $F^{\rm Brownian}_a$, but this is an artefact due to our definition of the Hariya-Yor process.  Indeed, under the scaling $u=at^{-1/3}, v=bt^{-1/3}$, $\lim_{t\to\infty} t^{-1/3} \mathcal{HY}_{u,v}(0)\overset{(d)}{=}E_{a+b}+E_{a-b}$, where $E_{a+b}$ and $E_{a-b}$ are independent exponential random variables with parameters $a\pm b$. We can construct a centered variable as follows: we may write that 
\begin{equation} \label{eq:defGab} 
    \lim_{t\to \infty} \mathbb P\left( \frac{h_{at^{-1/3}}(0,t)-h_{at^{-1/3}}(0,0)+\frac{t}{12} }{ t^{1/3} } \leqslant s \right)  = F^{\rm HY}_{a,b}(s)
\end{equation}
where $F^{\rm HY}_{a,b}(s)$ is such that 
\begin{equation}
    \mathbb E\left[ F^{\rm HY}_{a,b}(s- E_{a+b}-E_{a-b})\right] = G_{a,b}^{\rm HY}(s).
\end{equation}
More explicitly, this means that 
\begin{equation} \label{FtoG} 
    G^{\rm HY}_{a,b}(s) = (a^2-b^2) \int_{0}^{+\infty} \mathrm d u \,  e^{-ua } \frac{\sinh( bu)}{b}F_{a,b}^{\rm HY}(s-u).
\end{equation}
This can be inverted as  
\begin{equation}
F_{a,b}^{\rm HY}(s)= G_{a,b}^{\rm HY}(s)+ \partial_s G_{a,b}^{\rm HY}(s) \frac{2a}{a^2-b^2} + \partial^2_s G_{a,b}^{\rm HY}(s) \frac{1}{a^2-b^2}.
\end{equation}

Equivalently, if we denote by $\chi_{a,b}$ a random variable with distribution $G_{a,b}^{\rm HY}$ and $\xi_{a,b}$ a random variable with distribution $F^{\rm HY}_{a,b}$, we have that (recall that we have assumed in this section that  $b\leq 0$, $a+b>0$)
\begin{equation}
    \xi_{a,b}+E_{a+b}+E_{a-b}= \chi_{a,b}.
\end{equation}
where $\xi_{a,b}$, $E_{a+b}$ and $E_{a-b}$ are independent.
By stationarity, we can argue that $\mathbb E\left[ \xi_{a,b} \right]=0$. Indeed, 
\begin{align*}
    \mathbb E\left[ \xi_{a,b}\right] &=\lim_{t\to\infty} t^{-1/3} \mathbb E\left[ h_{at^{-1/3}}(0,t)-h_{at^{-1/3}}(0,0)+\frac{t}{12} \right] \\
    &= \lim_{t\to\infty} t^{-1/3} \int_{0}^t \left(\partial_s\mathbb E\left[ h_{at^{-1/3}}(0,s)-h_{at^{-1/3}}(0,0)\right] +\frac{1}{12}\right) \rmd s
\end{align*}
The quantity inside the integral is independent of $s$, by stationarity, so it is a constant. This constant is necessarily $0$, otherwise we would have that $\lim_{t\to\infty} \frac{h_{at^{-1/3}}(0,t)}{t}\neq \frac{-1}{12}$.  
%\end{remark}

\begin{remark} 
In view of Remark \ref{rem:specialcaseBrownian}, we should obtain that as $b\to -a$, 
\begin{equation} 
F^{\rm HY}_{a,b}(s) \xrightarrow[b\to -a]{}F^{\rm Brownian}_a(s),
\end{equation}
although this is not immediately apparent from our formulas. 
\end{remark}

\subsubsection{Summary}
\begin{figure}
\centering
    \begin{tikzpicture}[scale=1.5, every text node part/.style={align=center}]
    \draw[thick] (0,4) -- (4,0);
    \draw[thick] (0,0) -- (2,2);
    \draw[thick, ->] (2,2) -- (2,4.1) node[anchor=west] {$b$ \\\footnotesize   drift \\ \footnotesize  parameter};
    \draw[thick, ->] (2,2) -- (4.1,2) node[anchor=west] {$a$ \\ \footnotesize  boundary parameter};
    \draw[dashed, gray] (2,0) -- (2,2);
    \draw[dashed, gray] (0,2) -- (2,2);
    \draw[gray] (2,0.1) -- (2,-0.1) node[anchor=north] {$0$};
    \draw[gray] (0.1,2) -- (-0.1,2) node[anchor=east] {$0$};
    \draw (0.8,2.5) node{$F_a^{\rm Brownian}$};
    \draw (0.8,1.5) node{$F_a^{\rm Brownian}$};
    \draw (1.3,3.5) node{$F_a^{\rm Brownian}$};
    \draw (3.5,1.2) node{$F_{a,b}^{\rm HY}$};
    \draw (2.5,0.7) node{\bf ?};
    \draw (1.5,0.7) node{\bf ?};
    \draw (3,3) node{$F_{a,0}^{\rm HY}$};
    \draw[-stealth] (5,4) node{$F_a^{\rm Brownian}$} to[bend left] (2.5,1.5);
    \end{tikzpicture}    
    \caption{Definition of $F_{a,b}^{\rm stat}$ as in \eqref{eq:defFabstat} for various values of $a,b$. In a small portion of the phase diagram below the line $a+b=0$, we are not able to characterize the distribution of large time fluctuations.}
    \label{fig:cases}
\end{figure}
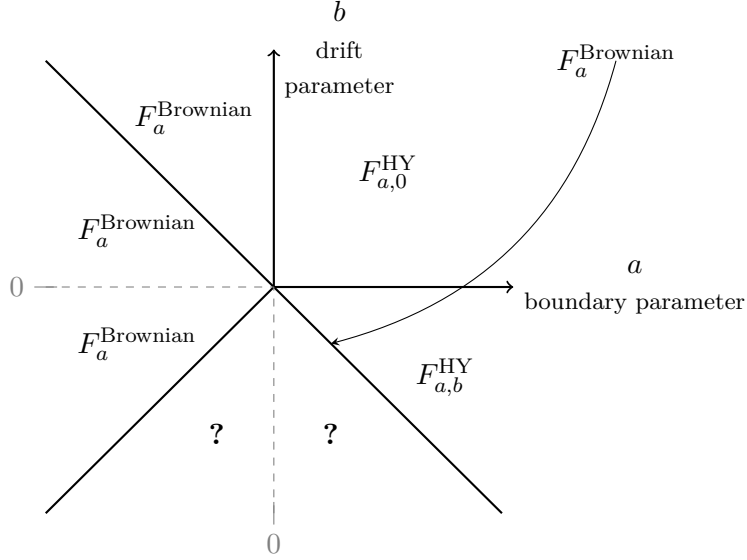

The results presented above for the different phases of the diagram in Fig. \ref{fig:half-space} can be summarized in a unified manner. We expect that for any $a,b \in \mathbb R$, scaling $u=at^{-1/3}, v=bt^{-1/3}$, when one considers the initial condition $h(x,0)=h^{\rm stat}_{u,v}(x)$ defined in \eqref{eq:defhstat}, we have that   
\begin{equation} \label{eq:generalconv} 
    \lim_{t\to \infty} \mathbb P\left( \frac{h_{at^{-1/3}}(0,t)+\frac{t}{12} }{ t^{1/3} } \leqslant s \right)  = F_{a,b}^{\rm stat}(s),
\end{equation}
for some CDF  $F_{a,b}^{\rm stat}$. We have obtained \eqref{eq:generalconv} and determined a formula for  $F_{a,b}^{\rm stat}$ in a number of cases (see Fig. \ref{fig:cases}), in particular:
\begin{equation}
\label{eq:defFabstat}
    F_{a,b}^{\rm stat}(s) = \begin{cases} 
    F^{\rm HY}_{a,0}(s)&\mbox{ for }a>0, b\geqslant 0,\\
    F_a^{\rm Brownian}&\mbox{ for }a\leqslant 0, b\geqslant a,\\
    F^{\rm HY}_{a,b}(s)&\mbox{ for } a>b, b\leqslant 0, a+b>0,\\
    F_a^{\rm Brownian}&\mbox{ for }a\geqslant 0, a+b=0,
    \end{cases}
\end{equation}
where $ F_a^{\rm Brownian}$ is defined in Section \ref{sec:introBrowniancase} and $F^{\rm HY}_{a,b}$ is defined in Section \ref{sec:introHYcase}.  In the next Sections, we will denote by $\xi_{a,b}$ a random variable with distribution $F_{a,b}^{\rm stat}$. 

\subsubsection{Away from the boundary} 

While we understand now the distribution and asymptotics of $h_u(0,t)$ and we know as well the distribution of stationary spatial increments $h_u(x,t)-h_u(0,t)$, we cannot immediately deduce the asymptotics of $h_u(x,t)$ for $x>0$. This is due to the fact that the random variables $h_u(0,t)$ and $h_u(x,t)-h_u(0,t)$ are correlated in a very nontrivial way. 

However, the distribution of $h_u(x,t)$ can be computed explicitly, at least in the large time limit, for the same initial condition as in the previous sections. When  $x=\tilde xt^{2/3}$, we obtain  
\begin{equation} 
\lim_{t\to \infty} \mathbb P\left( \frac{h_{at^{-1/3}}(t^{2/3}\tilde x,t)+\frac{t}{12} }{ t^{1/3} } \leqslant y\right)  = G_{a,b}^{\rm HY}(y, \tilde x). 
\end{equation}
The CDF is computed in Appendix \ref{sec:xpositif} and given in \eqref{eq:defGforxpositif}.

\subsection{Two-time covariance}
\label{sec:twotimeintro}

An interesting application of our results concerns the two-time covariance. 
For general initial conditions,  one can study the correlation of the KPZ height field at two different times, i.e. the correlation of $h(0,t_1)$ 
and $h(0,t)$ (here we focus on the same space point). In the limit where both times are large $1 \ll t_1 < t$ with a fixed ratio $\tau=t_1/t$, 
this covariance becomes a universal function $C(\tau)$ of $\tau \in [0,1]$, whose specific form depends only on the class of initial condition 
(stationary, droplet, flat, etc\dots ). This covariance, as well as the full two-time height distribution, were studied for the full-space geometry
\cite{ferrari2016time,de2017memory,de2017tail,le2017maximum,johansson2019two,de2018two,johansson2020long,liao2022multi, ferrari2019time, corwin2021kpz} and measured in experiments \cite{takeuchi2012evidence}. One defines for a general initial condition, 
\begin{equation} 
C(\tau) 
=\lim_{t \to +\infty}  \frac{1}{ t^{2/3}} {\rm Cov}(h_u(0,t \tau), h_u(0,t))
\end{equation}
The results of this paper allow to obtain some predictions for $C(\tau)$ in the half-space geometry.
In particular we obtain $C(\tau)$ for the stationary initial conditions for all $0<\tau<1$,
and for droplet initial conditions for close times, i.e. $1-\tau \ll 1$. Of particular 
interest are the dimensionless ratios 
\begin{align}  \label{eq:defrhoR} 
\rho(\tau) &= \lim_{t \to +\infty} \frac{{\rm Cov}(h_u(0,t \tau), h_u(0,t))}{\sqrt{{\rm Var} h_u(0,t \tau) {\rm Var} h_u(0,t)}   } 
= \frac{C(\tau)}{\tau^{1/3} C(1)},\\ 
R(\tau) &= \lim_{t \to +\infty} \frac{{\rm Cov}(h_u(0,t \tau), h_u(0,t))}{{\rm Var} h_u(0,t \tau)}  = \frac{C(\tau)}{\tau^{2/3} C(1)}, 
\end{align}
which measure the overlap of the two polymer configurations (of lengths $\tau t$ and $t$ respectively). 
In particular a finite value of $R(0)$, i.e. in the limit of infinitely separated times is a
measure of memory or ergodicity breaking \cite{takeuchi2012evidence,de2017memory,le2017maximum}. 

Let us first consider the droplet initial condition, and start by recalling 
the result for the full-space problem. 
For close times, i.e. in the limit $\tau\to 1$, one expects that the height profile reaches local stationarity at the intermediate time, which allows to compute the two-time covariance. Indeed it was shown that \cite{ferrari2016time} 
\begin{equation}
    C^{\rm droplet}(\tau) = \Var[\xi_{\rm GUE}] -\frac 1 2 \Var[\xi_{\rm BR}](1-\tau)^{2/3}+ \mathcal O(1-\tau),
\end{equation}
where $\xi_{\rm GUE}$ follows the GUE Tracy-Widom distribution and $\xi_{\rm BR}$ denotes a random variable following the Baik-Rains distribution \cite{baik2000limiting}. We have  $\Var[\xi_{\rm BR}] \approx 1.1504$. In this paper we extend the arguments of \cite{ferrari2016time} to the half-line geometry.
The results depend on the value of the boundary parameter $u$.
In the critical region ($u$ close to zero) it is natural to scale the boundary parameter $u= a t^{-1/3}$, and, as shown below, we obtain for the droplet initial condition (centered at $x=0$), as $\tau\to 1$, 
\begin{equation}
    C^{\rm droplet}(\tau) = \Var\left[ \xi_a^{\rm droplet} \right]  - \frac{1}{2}  \Var\left[\xi_{0,0}\right] (1-\tau)^{2/3} + \mathcal O(1-\tau),
    \label{eq:introcorrelationdroplet}
\end{equation}
where $\xi_{a}^{\rm droplet}$ denotes a random variable with distribution $F^{\rm droplet}_a$ defined in \eqref{eq:Fdroplet}, and $\xi_{0,0}$ denotes a random variable with distribution $F^{\rm stat}_{0,0} = F^{\rm Brownian}_0$. The variance was computed in \cite{barraquand2020half}, we have  $\Var[\xi_{0,0}] \approx 1.649$.
In the particular case $a=0$ the variable $\xi_0^{\rm droplet}$ has the GOE Tracy Widom distribution and its 
variance is $\Var[\xi_0^{\rm droplet}] \approx 1.6078$. In the limit $a \to +\infty$ one enters the unbound phase
and $\xi_{\infty}^{\rm droplet}$ has the GSE Tracy-Widom distribution \cite{tracy1996orthogonal} with
variance is $\Var[\xi_\infty^{\rm droplet}] \approx 0.5177$.

\bigskip 
Consider now stationary initial conditions. In that case one can compute $C(\tau)$ for arbitrary $\tau \in [0,1]$. Let us recall the result in full-space,
obtained in \cite{ferrari2016time}. If the initial condition is given by a standard Brownian motion, one obtains  
\begin{equation}
C(\tau) = \frac{1}{2} \left(1+\tau^{2/3} -(1-\tau)^{2/3}\right) \Var[\xi_{\rm BR}].
\end{equation} 
Note that it leads to the dimensionless ratio $R(\tau) = \frac{C(\tau)}{\tau^{2/3}} \to_{\tau \to 0} \frac{1}{2}$
which shows a memory effect at very separated times. We have extended the arguments of \cite{ferrari2016time} 
to the half-line geometry. The results depend on the value of the boundary and drift parameters $u,v$.
Let us start with the discussion of the critical region with $u=a t^{-1/3}$, $v=b t^{-1/3}$. We find, for any $a,b$, starting from the initial condition $h^{\rm stat}_{at^{-1/3}, bt^{-1/3}}$, 
\begin{equation}
C(\tau) = \frac{1}{2} \left( \Var (\xi_{a,b}) + \tau^{2/3} \Var (\xi_{a\tau^{1/3}, b\tau^{1/3}}) - (1-\tau)^{2/3}\Var (\xi_{a(1-\tau)^{1/3},b(1-\tau)^{1/3}})\right).
\end{equation}
where $\xi_{a,b}$ a random variable with distribution $F_{a,b}^{\rm stat}$ defined in \eqref{eq:defFabstat}. 

\begin{remark}
For $\tau \to 0$ one finds $R(0^+)= \frac{1}{2} \frac{ \Var (\xi_{0,0}) }{ \Var (\xi_{a,b})}$ 
\end{remark} 

\begin{remark}  Consider the case  $a=0$ and $b\geqslant 0$. For $b\geqslant 0$, $\xi_{a,b}=\xi_{a,0}$ by definition, and for $a=0$, $\xi_{a,0}=\xi_{0,0}=\xi_0^{\rm Brownian}$, so that  the dimensionless ratio defined in \eqref{eq:defrhoR}
is asymptotically 
\begin{equation} 
\rho(\tau) = \frac{1}{2\tau^{1/3}}\left(1+\tau^{2/3}-(1-\tau)^{2/3} \right),
\label{eq:rhosimple}
\end{equation}
that is exactly the same as for the KPZ equation in full-space \cite{ferrari2016time}. 
For general $a,b$ this is not the case however. Using Remark \ref{rem:atoinfty}, the same formula is true in the case $a=+\infty, b\geqslant 0$ (since $\xi_{+\infty,0}=\xi_0^{\rm Brownian}$).
\label{rem:covariancesimple}
\end{remark} 

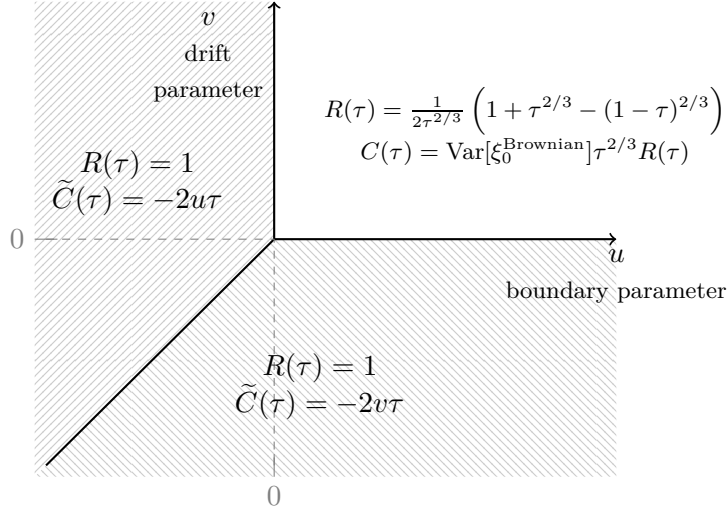
\begin{figure}
    \centering
    \begin{tikzpicture}[scale=1.5, every text node part/.style={align=center}]
    \fill[pattern=north east lines, pattern color=black!20] (-.1,-.1) --(2,2) -- (2,4.1) -- (-.1,4.1) -- cycle;
    \fill[pattern=north west lines, pattern color=black!20] (-.1,-.1) --(2,2) -- (5,2) -- (5,-.1) -- cycle;
    \draw[thick] (0,0) -- (2,2);
    \draw[thick, ->] (2,2) -- (2,4.1) node[anchor=north east] {$v$ \\\footnotesize   drift \\ \footnotesize  parameter};
    \draw[thick, ->] (2,2) -- (5,2) node[anchor=north] {$u$ \\ \footnotesize  boundary parameter};
    \draw[dashed, gray] (2,0) -- (2,2);
    \draw[dashed, gray] (0,2) -- (2,2);
    \draw[gray] (2,0.1) -- (2,-0.1) node[anchor=north] {$0$};
    \draw[gray] (0.1,2) -- (-0.1,2) node[anchor=east] {$0$};
    \draw (0.8,2.5) node{$R(\tau)=1$\\ $\widetilde C(\tau)=-2u\tau$};
    \draw (2.4,0.7) node{$R(\tau)=1$\\ $\widetilde C(\tau)=-2v\tau$};
    \draw[align=left] (4.2,3) node[align=left]{\footnotesize  $R(\tau) = \frac{1}{2\tau^{2/3}}\left(1+\tau^{2/3} -(1-\tau)^{2/3}\right)$ \\ \footnotesize  $C(\tau) = \Var[\xi_0^{\rm Brownian}] \tau^{2/3}  R(\tau)$};
    \end{tikzpicture}
    \caption{Phase diagram of two time covariances, for stationary initial condition $h^{\rm stat}_{u,v}$ with fixed $u,v$. } 
    \label{fig:covariances}
\end{figure}

Let us now discuss what happens for fixed $u,v$. There are three phases which are depicted in Fig. \ref{fig:covariances} 
where the main results are summarized. The first phase is defined by $u,v>0$ corresponding by to taking the limit $a,b\to+\infty$ in the previous discussion. We find that the two-time covariance does not depend on $u,v$ and is given by \eqref{eq:rhosimple} (indeed, as explained in Remark \ref{rem:covariancesimple}, $\xi_{+\infty, +\infty}= \xi_0^{\rm Brownian}$). 

If $u<0$ or $v<0$, however, the scalings will be different with height fluctuations of order $t^{1/2}$ instead of $t^{1/3}$.
We define a variant of the coefficient $C(\tau)$ by 
\begin{equation}
    \widetilde C(\tau) = \lim_{t \to +\infty}  \frac{1}{ t} {\rm Cov}(h_u(0,t \tau), h_u(0,t)).
\end{equation}
To discuss the phase $u<0$ (with $u<v$), we consider the equivalent polymer picture. In that phase, the polymer is bound to the wall (with Gaussian free energy fluctuations) so that $\mathrm{Cov}(h_u(0,t \tau), h_u(0,t)) \simeq \Var(h_u(0,t \tau))$ at large time, which is known \cite{barraquand2021kardar,deNardisPLDTT} to be asymptotically equivalent  to $-2ut\tau$. This implies that $\tilde C(\tau)= -2 u \tau$. 
The $R$ ratio being now equal to $R(\tau)= \frac{C(\tau)}{\tau C(1)}$ one finds that it is equal to $R(\tau)=1$ in that phase.

In the phase $v<0$ (with $v<u$), the covariance will be determined by the initial condition, which can be approximated at large scale by a Brownian motion with drift $-v$. The free energy equals  
\begin{multline} 
h_u(0,t) = \log Z(0,t) = \log\left( \int_0^{+\infty} dx Z(x,0 \vert 0,t)e^{h_u(x,0)} \right)\\ 
\approx \max_x \left( \log Z(x,0\vert 0,t) + h_u(x,0) \right) .\end{multline} 
The optimal $x=x_{\rm max}$ can be approximated by $\argmax \left\lbrace \frac{x^2}{4t} + vt\right\rbrace $ so that $x_{\rm max}\simeq -2v t$ and the fluctuations of the free energy $\log Z(x,0\vert 0,t)$ are subdominant compared to fluctuations of $h(x,0)$. 
Thus, we again have that $\mathrm{Cov}(h_u(0,t \tau), h_u(0,t)) \simeq \Var(h_u(0,t \tau))$ which is asymptotically equivalent to the variance of the initial condition $\Var(h_u(x,0))$ at the point $x=-t\tau v$. Since the initial condition can be approximated at large scale by a Brownian motion with drift $-v$, we find that  $\mathrm{Cov}(h_u(0,t \tau), h(0,t)) = -2vt\tau$. This implies that $\tilde C(\tau)= -2 v \tau$, and again the $R$ ratio is again equal to $R(\tau)= \frac{C(\tau)}{\tau C(1)}$ one finds that it is equal to $R(\tau)=1$ in that phase.

\subsection{Mathematical aspects}

Let us stress a few points that deserve further consideration from a mathematical perspective. First of all, outside of the Brownian phase  ($a\leq 0$, $a\leq b$), the fact that the Hariya-Yor processes defined in \eqref{eq:defhstat} are stationary for the half-space KPZ equation was discovered in \cite{barraquand2021steady}. These processes arose as $L\to +\infty$ limits of stationary processes for the KPZ equation on $[0,L]$, for which formulas had been found in \cite{corwin2021stationary}. That the $L\to+\infty$ limit of stationary processes on $[0,L]$ are stationary for the dynamics on $\mathbb R_+$ is a very reasonable hypothesis, but it still needs to be formally proven mathematically.

The computation of limiting distributions obtained in this article rely on a combination of physics and mathematics methods, but we focus in this article on physics results and do not attempt to prove the results according to the standards of writing in the mathematics literature.  We refer to \cite[Section 2.4]{barraquand2020half} where we had already discussed the interplay between these physics and mathematics arguments, as well as the main challenges that would arise to turn these results into mathematical proofs. The results of \cite[Section 10]{imamura2021skew} would likely be useful in order to prove rigorously the Pfaffian formula for the generating series \eqref{eq:5.14}.
 
The results of Section \ref{sec:twotimeintro} also rely on a mixture  of established facts, and  assumptions based on analogies with the full-space setting, about the universal processes describing the large time fluctuations of $h(x,t)$, starting from various initial conditions. It would be interesting to confirm those assumptions by rigorous proofs.

\subsubsection*{Acknowledgements} This article is based upon work supported by the National Science Foundation under Grant No. DMS-1928930 while the three authors participated in a program hosted by the Mathematical Sciences Research Institute in Berkeley, California, during the Fall 2021 semester. G.B. was partially supported by the ANR grant CORTIPOM. AK acknowledges support from ERC under Consolidator grant number
771536 (NEMO). PLD acknowledges support from ANR grant ANR-17-CE30-0027-01 RaMaTraF.

\section{Moment formula}
\label{sec:moments}
\subsection{Nested contour moment formula}

 In order to study this initial condition \eqref{eq:defHariyaYor}, we will first study formulas from the more general initial condition 
  \begin{equation}
    \mathcal{Z}_{u,v_1,v_2}(x)
    = w_1  \int_{0}^x\mathrm d t e^{B_1(t)+B_2(x)-B_2(t)} + w_1w_2 e^{B_2(x)},
     \label{eq:defHariyaYorgeneralized}
 \end{equation}
 where $B_1, B_2$ are independent standard Brownian motions with drifts $-v_1$ and $-v_2$ respectively, and $w_1, w_2$ are independent inverse gamma random variables \break $w_1\sim\mathrm{Gamma}^{-1}(u+v_1)$ and $w_2\sim\mathrm{Gamma}^{-1}(u+v_2)$. Equation \eqref{eq:defHariyaYorgeneralized} can be interpreted as a two-row O'Connell-Yor semi-discrete polymer partition function with inverse gamma weights in the beginning of each row. Eventually, we will let $v_2=-v_1=v$, so that $\log(\mathcal{Z}_{u,v_1,v_2}(x)) = \mathcal{HY}_{u,v}(x)$.
 \begin{remark} 
 Similar partition functions of O'Connell-Yor type polymers with inverse Gamma decorations have been  considered earlier in  \cite{borodin2015height, talyigas2020borodin} in the context of full-space KPZ growth and in \cite{barraquand2021kardar} in the context of KPZ equation in a half-space. 
 \end{remark}

\begin{figure}
    \centering
    \begin{tikzpicture}[scale=1.3]
    \draw[ultra thick, -stealth] (-2,0) -- (8,0);
    \draw[ultra thick, -stealth] (0,-2) -- (0,3); 
    \draw (-0.2,-0.2) node{$0$};
    \draw (5.5,3) -- (5.5,-2) node[anchor=north] {\footnotesize $r_1+\I\R$};
    \draw[-stealth] (5.5,-2) -- (5.5,2);
    \draw (4.5,3) -- (4.5,-2) node[anchor=north] {\footnotesize $r_2+\I\R$};
    \draw[-stealth] (4.5,-2) -- (4.5,2);
    \draw (3.5,3) -- (3.5,-2) node[anchor=north] {\footnotesize $r_3+\I\R$};
    \draw[-stealth] (3.5,-2) -- (3.5,2);
    \draw (2.5,1) node{...};
    \draw (1.5,1) node{...};
    \draw (0.5,3) -- (0.5,-2) node[anchor=north] {\footnotesize $r_k+\I\R$};
    \draw[-stealth] (0.5,-2) -- (0.5,2);
    \draw (6.5,0.1) -- (6.5,-0.1) node[anchor=north]{\footnotesize $v_1-\tfrac 1 2$};
    \draw (7,-0.1) -- (7,0.1) node[anchor=south]{\footnotesize $v_2-\tfrac 1 2$};
    \draw (-1,-0.1) -- (-1,0.1) node[anchor=south]{\footnotesize $\tfrac 1 2 -u$};
    \draw[stealth-stealth] (4.6,2.5) -- (5.4,2.5);
    \draw (5,2.8) node{$>1$};
    \end{tikzpicture}
    \caption{The contours used in \eqref{eq:defv}.}
    \label{fig:nestedcontours}
\end{figure}
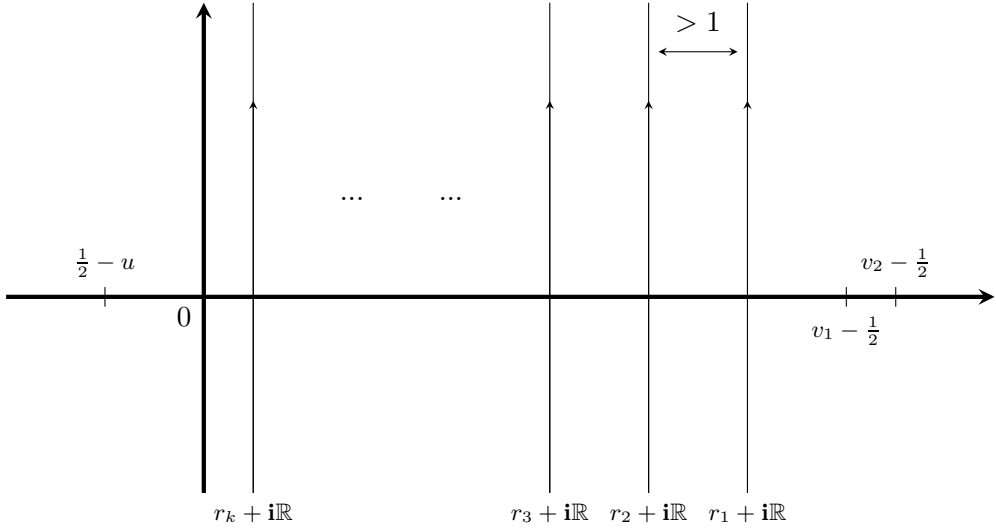

For $x_1\geqslant x_2\geqslant \dots \geqslant x_k\geqslant 0$, define  $f(t,\vec x) = \mathbb E\left[ Z(x_1,t)\dots Z(x_k,t)\right]$ where we assume the initial condition  $Z(x,0) = \mathcal{Z}_{u,v_1,v_2}(x)$. The function $f(t,\vec x)$  satisfies the following conditions \cite{borodin2016directed} (see also \cite{kardar1987replica, gaudin1983fonction}). It satisfies the heat equation
 	\begin{equation}
 	\partial_t f(\vec x,t) = \sum_{i=1}^k \partial_{x_i}^2 f(t,\vec x),
 	\label{eq:heat}
 	\end{equation}
 	on the sector $x_1\geqslant x_2\geqslant \dots \geqslant x_k\geqslant 0$,  subject to the two-body boundary condition  
 	\begin{equation}
 	\left( \partial_{x_{i+1}}-\partial_{x_i}-1\right)f \big\vert_{x_i=x_{i+1}}=0,
 	\label{eq:twobodyboundary}
 	\end{equation}
 with a boundary condition at $0$ given by  
 	\begin{equation}
 	\left(\partial_{x_k}-(u-\frac 1 2)\right)f\big\vert_{x_k=0} = 0. 
 	\label{eq:boundary0}
 	\end{equation}
 The function $f(t,\vec x)$ must also satisfy  the initial condition
 	\begin{equation}
 	f(\vec x,t=0) = \mathbb E\left[\prod_{i=1}^k Z(x_i,0)\right].
 	\label{eq:initialcondition}
 	\end{equation}
 	Fix $k\geqslant 1$ and assume that $u,v_1,v_2\in \mathbb R_{\geqslant 0}$ are such that  $v_1-\frac{1}{2}>k-1, v_2-\frac{1}{2}>k-1$ (note that this hypothesis is necessary and it was missing in \cite[Claim 4.7]{barraquand2020half}) and $u+v_1, u+v_2>k$. 
The function 
	\begin{multline}
f(\vec x, t \vert u,v_1,v_2) = \frac{2^k \Gamma(v_1+v_2)}{\Gamma(v_1+v_2-k)} 
\int_{r_1+\I\R}\frac{\mathrm{d}z_1}{2\I\pi}\cdots \int_{r_k+\I\R}\frac{\mathrm{d}z_k}{2\I\pi} \prod_{1\leqslant a<b\leqslant k} \frac{z_a-z_b}{z_a-z_b-1}\, \frac{z_a+z_b}{z_a+z_b-1}\\ \times
\prod_{i=1}^k \frac{z_i}{z_i+u-\frac 1 2}  \frac{1}{(v_1-1/2)^2-z_i^2}\frac{1}{(v_2-1/2)^2-z_i^2}  e^{tz_i^2 - x_i z_i},
\label{eq:defv}
\end{multline}
where the contours (see Fig. \ref{fig:nestedcontours}) are chosen so that 
$$\min\left\lbrace v_1-\frac 1 2, v_2-\frac 1 2 \right\rbrace >r_1>r_2+1>\dots, > r_k+k-1>\max\lbrace k-1-u+\frac 1 2, k-1\rbrace,$$ 
 satisfies the equations \eqref{eq:heat}, \eqref{eq:twobodyboundary}, \eqref{eq:boundary0} and the initial condition \eqref{eq:initialcondition} with $Z_0(x) = \mathcal{Z}_{u,v_1,v_2}(x)$ defined in \eqref{eq:defHariyaYorgeneralized}. 
Assuming that there is at most one solution to these equations, we obtain that for $x_1\geqslant x_2\geqslant \dots \geqslant x_k\geqslant 0$, $u+v_1>k$, $u+v_2>k$, $ v_1-1/2>k-1$ and $ v_2-1/2>k-1$, 
\begin{equation} 
\label{eq:momentformula}
\mathbb E\left[ Z(x_1,t)\dots Z(x_k,t)\right] = f(\vec x, t \vert u,v_1,v_2)
\end{equation}
where $f(\vec x, t \vert u,v_1,v_2)$ is defined in \eqref{eq:defv}. 

\begin{remark}
When  $v_1 \to +\infty$ one has that $Z_0(x) \to w_1 w_2 e^{B_2(x)}$ and thus one 
recovers the result (4.19) in \cite{barraquand2020half}, taking into account that $\overline{w_{1}^k} = \frac{\Gamma(v_1+u-k)}{\Gamma(v_1+u)}$ and $\overline{w_{2}^k} = \frac{\Gamma(v_2+u-k)}{\Gamma(v_2+u)}$.
When  $v_2 \to +\infty$ one has $v_2 Z_0(x) \to w_1 e^{B_1(x)}$ and the same works.
\end{remark}

Let us now explain why $f(\vec x, t \vert u,v_1,v_2)$ satisfies  each of the equations \eqref{eq:heat}, \eqref{eq:twobodyboundary}, \eqref{eq:boundary0} and \eqref{eq:initialcondition} one by one (except for \eqref{eq:initialcondition}, the arguments are based on \cite{borodin2016directed} and \cite{borodin2014macdonald}). 
\medskip 

 The function $\prod_{i=1}^k  e^{tz_i^2 - x_i z_i}$ is a solution of \eqref{eq:heat} for any $\vec z$, so that by linearity, the function  $f(\vec x, t \vert u,v_1,v_2)$ is also a solution. 

\medskip 

 Let us apply the operator $\left( \partial_{x_{i+1}}-\partial_{x_i}-1\right)$  to $f(\vec x, t \vert u,v)$ and let us assume that $x_i=x_{i+1}$ for some $i$. The application of the operator brings a factor $(-z_{i+1}+z_i-1)$ inside the integrand in \eqref{eq:defv}. This extra factor cancels with the denominator $z_a-z_b-1$ for $a=i, b=i+1$, so that there is no pole anymore at $z_{i+1}=z_{i}-1$. Hence, we may deform the contour for $z_{i+1}$ to be the same as the one for $z_i$. Because of the factor $z_a-z_b$   for $a=i, b=i+1$, the integrand is now antisymmetric with respect to exchanging $z_i$ and $z_{i+1}$, and since both variables are integrated along the same contour, their integral is zero.  We conclude that $f(\vec x, t \vert u,v)$ satisfies  \eqref{eq:twobodyboundary}.

\medskip 

 Let as apply the operator $\left(\partial_{x_k}-u+\frac 1 2\right)$ to $f(\vec x, t \vert u,v)$. This brings an extra factor $-z_k-u+\frac 1 2 $ to the integrand, which cancels the denominator $z_k+u-\frac 1 2$ already present in the formula. After this cancellation, and if one assumes that $x_k=0$, then the integrand is antisymmetric with respect to changing $z_k$ into $-z_k$.  Furthermore, the integrand has no pole anymore at $z_k=-(u-\frac 1 2)$ so that the contour of $z_k$ can be freely deformed to the left, regardless of the value of $u$. Since $v-\frac 1 2>k-1$, it is then always possible to shift the $z_k$ contour to the left so that  $v-\frac 1 2>r_1>r_2+1 > \dots > r_{k-1}+k-2>\max\lbrace k-1, k-2-(u-\frac 1 2)\rbrace$ and $r_k=0$. Now, the integration contour for $z_k$ is symmetric with respect to changing $z_k$ into $-z_k$, and thus the integral is zero. We conclude that $f(\vec x, t \vert A,B)$ satisfies  \eqref{eq:boundary0}.
\medskip 

 Let us assume that $t=0$ and $x_1>x_2> \dots x_k>0$. The formula agrees with moments of $\mathcal{Z}_{u,v_1,v_2}(x)$ using a scaling limit of the half-space log-gamma polymer moment formula from  \cite{barraquand2020halfMacdonald} (see \cite[Proposition 4.2]{barraquand2020half}). More precisely, we take $\alpha_{\circ}=u, \alpha_1=v_1, \alpha_2=v_2$ and for $i\geqslant 3$, $\alpha_i=\frac{1}{2}+\sqrt{n}$ and consider the log-gamma partition function $\mathcal Z(\sqrt{n}x/2,2)$. The scaling limit is explained in \cite[Section 4.3]{barraquand2020half}, see also \cite[Appendix C.3]{barraquand2021kardar}.

 \subsection{Pfaffian formula}
\label{sec:Pfaffian}

Let us define the two functions 
\begin{equation}
\mathtt{G}(z) = \frac{z^3}{3}-\frac{z^2}{2}+\frac{z}{6}, \qquad \ratioGamma(z) = \frac{\Gamma(u-z)}{\Gamma(u+z)}\frac{\Gamma(v_1-z)}{\Gamma(v_1+z)}\frac{\Gamma(v_2-z)}{\Gamma(v_2+z)} \Gamma(2z)
\label{GG} 
\end{equation}

Then, for $u-\frac{1}{2}, v_1-\frac 1 2 , v_2-\frac 1 2 >k-1$, using the moment formula \eqref{eq:momentformula} and similar manipulations 
as in \cite[Sections 4.4 and 4.7]{barraquand2020half}, based on \cite[Conjecture 5.2]{borodin2016directed}, we obtain the expression for the moment of  $Z(0,t)$ as
\begin{equation}
\begin{split}
\EE[Z(0,t)^k] &= 4^k k! \frac{\Gamma(v_1+v_2)}{\Gamma(v_1+v_2-k)} \sum_{\underset{\lambda=1^{m_1}2^{m_2}\dots}{\lambda\vdash k}} \frac{(-1)^{\ell(\la)}}{m_1!m_2!\dots} \int_{\I\R} \frac{\rmd w_1}{2\I\pi} \dots \int_{\I\R} \frac{\rmd w_{\ell(\la)}}{2\I\pi}\\
&\times \Pf \left[ \frac{u_i-u_j}{u_i+u_j} \right]_{i,j=1}^{2\ell(\la)}  \prod_{j=1}^{\ell(\la)}  \frac{e^{t \mathtt{G}(w_j+\la_j)}}{e^{t \mathtt{G}(w_j)}}   \frac{(w_j+1/2)_{\la_j-1}}{4(w_j)_{\la_j}} \frac{\Gamma(-w_j+1)\Gamma(w_j+\lambda_j)}{\Gamma(-w_j-\lambda_j+1)\Gamma(w_j)} \\
&\times \frac{\Gamma(u+\frac{1}{2}-w_j-\la_j)\Gamma(v_1+\frac{1}{2}-w_j-\la_j)\Gamma(v_2+\frac{1}{2}-w_j-\la_j)}{\Gamma(u-\frac{1}{2}+w_j+\la_j)\Gamma(v_1-\frac{1}{2}+w_j+\la_j)\Gamma(v_2-\frac{1}{2}+w_j+\la_j)}\\
&\times \frac{\Gamma(u-\frac{1}{2}+w_j)\Gamma(v_1-\frac{1}{2}+w_j)\Gamma(v_2-\frac{1}{2}+w_j)}{\Gamma(u+\frac{1}{2}-w_j)\Gamma(v_1+\frac{1}{2}-w_j)\Gamma(v_2+\frac{1}{2}-
w_j)},
\label{eq:conjecture} 
\end{split}
\end{equation}
where $(u_1, \dots, u_{2\ell(\la)}) = (-w_1+\frac 1 2 , w_1-\frac 1 2 +\lambda_1, \dots, -w_{\ell(\la)}+\frac 1 2 , w_{\ell(\la)}-\frac 1 2 +\lambda_{\ell(\la)})$ and the sum runs over integer partitions $\lambda$ of $k$. This can be rewritten as
\begin{multline}
\EE[Z(0,t)^k] = 4^k k! \frac{\Gamma(v_1+v_2)}{\Gamma(v_1+v_2-k)} \sum_{\underset{\lambda=1^{m_1}2^{m_2}\dots}{\lambda\vdash k}} \frac{(-1)^{\ell(\la)}}{m_1!m_2!\dots} \int_{\I\R} \frac{\rmd w_1}{2\I\pi} \dots \int_{\I\R} \frac{\rmd w_{\ell(\la)}}{2\I\pi}\\
\times \Pf \left[ \frac{u_i-u_j}{u_i+u_j} \right]_{i,j=1}^{2\ell(\la)}  \prod_{j=1}^{\ell(\la)}  \frac{e^{t \mathtt{G}(w_j+\la_j)}}{e^{t \mathtt{G}(w_j)}}   \frac{(w_j+1/2)_{\la_j-1}}{4(w_j)_{\la_j}} \frac{\Gamma(-w_j+1)\Gamma(w_j+\lambda_j)}{\Gamma(-w_j-\lambda_j+1)\Gamma(w_j)} \\
\times \frac{\ratioGamma(w_j-\frac{1}{2}+\la_j)\ratioGamma(-w_j+\frac{1}{2}) }{\Gamma(2w_j-1+2\la_j)\Gamma(1-2w_j)},
\label{eq:conjecture2} 
\end{multline}

We recognize the same formula as \cite[Claim 4.11, Eq. (4.35)]{barraquand2020half} with the only difference that the ratio $\ratioGamma(z)$ here should be replaced by 
\begin{equation}
G(z) =\frac{\Gamma(A+1/2-z)}{\Gamma(A+1/2+z)}\frac{\Gamma(B+1/2-z)}{\Gamma(B+1/2+z)} \Gamma(2z).
\end{equation}

\subsection{Moment series in terms of a Fredholm Pfaffian}
\label{sec:FP} 

We will now write the moment generating function of $Z(0,t)$. Let $W$ be a inverse gamma random variable with parameter $v_1+v_2$, independent from the initial condition and from the noise $\xi$. We define, for $\varsigma>0$, 
\begin{equation}
g(\varsigma) =\mathbb{E} \left[ \exp(- \varsigma e^{\frac{t}{12}} \invgamma Z(0,t)) \right].
\end{equation}
Ignoring the fact that the summation over $k$ cannot be exchanged with the expectation due to the divergence of moments, we will consider the following formal power series 
\begin{equation}
   1 + \sum_{k=1}^\infty \frac{(- \varsigma e^{\frac{t}{12}})^k}{k!} \mathbb{E} \left[\invgamma^k Z(0,t)^k\right],  
\end{equation}
that we will again denote by $g(\varsigma)$. This generating series was computed in \cite[Section 5]{barraquand2020half}, and leads to the Fredholm Pfaffian formula \cite[(5.16)]{barraquand2020half} with  the kernel \cite[(2.12)]{barraquand2020half}. Since our moment formula has exactly the same form with a different choice of $\ratioGamma$, we apply the same manipulations (it suffices to replace $\ratioGamma(z)$ in the present paper by $G(z)$ in \cite{barraquand2020half}) and obtain the following.  
\begin{equation}
g(\varsigma)=1+\sum_{\ell=1}^\infty\frac{(-1)^{\ell}}{\ell!} \prod_{p=1}^{\ell} \int_\mathbb{R} \rmd r_p\frac{\varsigma}{\varsigma + e^{-r_p}} {\rm Pf}\left[ K(r_i,r_{j})\right]_{i,j=1}^{2\ell(\la)} 
\label{eq:5.14}
\end{equation}
This series is a Fredholm Pfaffian,
\begin{equation}\label{eq:PfaffAllTimesAllA}
g(\varsigma)=\mathbb{E} \left[ \exp(- \varsigma e^{\frac{t}{12}} \invgamma Z(0,t)) \right]={\rm Pf}(J-\sigma_\varsigma K)_{\mathbb{L}^2(\mathbb{R})}. 
\end{equation}
The kernel $K$ is matrix valued and represented by a $2\times 2$ block matrix with elements
\begin{equation}\label{eq:kernelAllA}
\begin{split}
&K_{11}(r,r')=\iint_{C^2} \frac{\mathrm{d}w}{2\I\pi}\frac{\mathrm{d}z}{2\I\pi}\frac{w-z}{w+z}\ratioGamma(w)\ratioGamma(z)\cos(\pi w)\cos(\pi z)e^{ -rw-r'z + t  \frac{w^3+z^3}{3} },\\
&K_{22}(r,r')=\iint_{C^2} \frac{\mathrm{d}w}{2\I\pi}\frac{\mathrm{d}z}{2\I\pi}\frac{w-z}{w+z}\ratioGamma(w)\ratioGamma(z)\frac{\sin(\pi w)}{\pi}\frac{\sin(\pi z)}{\pi}e^{ -rw-r'z + t  \frac{w^3+z^3}{3}},\\
&K_{12}(r,r')=\iint_{C^2}  \frac{\mathrm{d}w}{2\I\pi}\frac{\mathrm{d}z}{2\I\pi}\frac{w-z}{w+z}\ratioGamma(w)\ratioGamma(z)\cos(\pi w)\frac{\sin(\pi z)}{\pi}e^{-rw-r'z + t  \frac{w^3+z^3}{3}},\\
&K_{21}(r,r')=-K_{12}(r',r).
\end{split}
\end{equation}
where the dependence in parameters $u,v_1,v_2$ only appears in the function $\ratioGamma(z)$ which was defined in \eqref{GG},
and the contour $C$ is an upwardly oriented vertical line parallel to the imaginary axis with real part between $0$ and $\min\lbrace u, v_1,v_2,1\rbrace $. The function $\sigma_\varsigma$ is given by $\sigma_\varsigma(r)=\frac{\varsigma}{\varsigma+e^{-r}}$
and the $2\times 2$ symplectic kernel $J$ is given by  $J(r,r')=\bigg(\begin{array}{cc}
0 & 1 \\ 
-1 & 0
\end{array} \bigg)\mathds{1}_{r=r'}$.

As in \cite[Section 2.2.2]{barraquand2020half} we may also rewrite $g(\varsigma)$ as the square root of a Fredholm determinant with a scalar kernel. We obtain 
\begin{equation} \label{res-scalar} 
g(\varsigma)=\mathbb{E} \left[ \exp(- \varsigma e^{\frac{t}{12}} \invgamma Z(0,t)) \right]={\rm Pf}(J-\sigma_\varsigma K)_{\mathbb{L}^2(\mathbb{R})}=\sqrt{\mathrm{Det}(I-\bar{K}_{t,\varsigma})_{\mathbb{L}^2(\mathbb{R}_+)}}.
\end{equation}
where the kernel  $\bar{K}_{t,\varsigma}$ is defined by 
\begin{equation}\label{eq:1DkernelAllT}
\bar{K}_{t,\varsigma}(x,y)={2\partial_x \iint_{C^2} \frac{\rmd w \rmd z}{(2\I\pi)^2 }\ratioGamma(z)\ratioGamma(w) \frac{\sin(\pi (z-w))}{\sin(\pi(z+w))}\varsigma^{w+z}e^{-xz-yw  + t  \frac{w^3+z^3}{3} } }
\end{equation}
From the generating function $g(\varsigma)$, one can compute the Laplace transform 
$$\mathbb E\left[ \exp(- \varsigma e^{\frac{t}{12}} Z(0,t)) \right]$$
using \cite[Remark 5.1]{barraquand2020half}, or perform a analogue of Laplace inversion using \cite[Remark 5.2]{barraquand2020half}. In the sequel, we will only be interested in the large time limit, in which case, asymptotics can be directly extracted from the asymptotics of $g(\varsigma)$.

\section{Large time  height distribution at $x=0$}
\label{sec:largetime}
\subsection{Limiting distribution in terms of a Fredholm determinant}
\label{subsec:largetime} 

As we have found in \cite{barraquand2020half}, at large times, 
\begin{equation}
    g(e^{- t^{1/3} s})  \simeq  \mathbb{P}\left(\frac{h_u(0,t)+ \log W + \frac{t}{12}}{t^{1/3}} \leq s\right).
\end{equation}
We will also rescale parameters as 
\be
u=at^{-1/3}, v_1=bt^{-1/3}, v_2=ct^{-1/3}.
\ee 
Under this scaling, $\log(W)/t^{1/3}$ converges to an exponential distribution independent from $h_{at^{-1/3}}(0,t)$, so that \cite[Eq. (4.3)]{ferrari2006scaling}
\begin{equation}
    \lim_{t \to +\infty} \mathbb{P}\left(\frac{h_{at^{-1/3}}(0,t)+\frac{t}{12} }{t^{1/3}}\leqslant s\right) = \left(1+\frac{\partial_s}{\Brescaled+\Crescaled} \right) \lim_{t\to\infty} g(e^{- t^{1/3} s}).
\end{equation}
To compute the limit of $g(e^{- t^{1/3} s})$, we use \eqref{eq:1DkernelAllT} with the scalings $(w,z) \to t^{-1/3} (w,z)$. 
The kernel $t^{1/3}\bar{K}_{t,e^{-t^{1/3}s}}(xt^{1/3}, yt^{1/3})$ converges to $\bar{K}^{(\Arescaled, \Brescaled, \Crescaled)}(x+s,y+s)$ where 
		\begin{equation}
	\bar{K}^{(\Arescaled, \Brescaled, \Crescaled)}(x,y) =\frac{1}{2}  \iint_{\Gamma_A^2} \frac{\rmd w \rmd z}{(2\I\pi)^2 }\ratioGammaLargeTime(w)\ratioGammaLargeTime(z)\frac{w-z}{w+z}\frac{1}{w} e^{-xz-yw +   \frac{w^3+z^3}{3} },
	\label{eq:Kbarscalaire}
	\end{equation}
 where the contour $\Gamma_A$ is an upwardly oriented vertical line with real part between $0$ and $\min\lbrace\Arescaled, \Brescaled, \Crescaled \rbrace$ as previously, and 
  \be \label{GGLargeTime} 
\ratioGammaLargeTime(z) = \frac{\Arescaled+z}{\Arescaled-z}\frac{\Brescaled+z}{\Brescaled-z}\frac{\Crescaled+z}{\Crescaled-z}.
\ee 
Thus, for any $\Arescaled, \Brescaled, \Crescaled >0$, 
 \begin{align} 
   G^{(\Arescaled, \Brescaled, \Crescaled)}(s) &:= \lim_{t \to +\infty} \mathbb{P}\left(\frac{h(0,t)+\frac{t}{12} }{t^{1/3}}\leqslant s\right)\label{eq:limproba}\\
   &=\left(1+\frac{\partial_s}{\Brescaled+\Crescaled} \right)\sqrt{\det(I-\bar K^{(\Arescaled, \Brescaled, \Crescaled)})_{\mathbb{L}^2(s,+\infty)}},  \label{Feq}
 \end{align}
where $\bar K^{(\Arescaled,\Brescaled, \Crescaled)}$ is defined in \eqref{eq:Kbarscalaire}.  Using the decomposition $\frac{1}{2} \frac{w-z}{(w+z)w} = \frac{1}{w+z} -\frac{1}{2w}$, 
	we obtain that 
	\begin{equation}
\bar K^{(\Arescaled, \Brescaled,  \Crescaled)}(x,y) = \int_{0}^{+\infty} \mathrm d \la A^{(\Arescaled, \Brescaled, \Crescaled)}(x+\la)A^{(\Arescaled, \Brescaled, \Crescaled)}(\la+y)\, -\frac 1 2 A^{(\Arescaled, \Brescaled, \Crescaled)}(x) \int_{0}^{+\infty}A^{(\Arescaled, \Brescaled, \Crescaled)}(y+\la)\, \mathrm d\la,
\label{eq:GLDform}
	\end{equation}
	where the function $A^{(\Arescaled, \Brescaled, \Crescaled)}(x)$ is defined by 
\begin{equation}
A^{(\Arescaled, \Brescaled, \Crescaled)}(x) = \int_{\Gamma_A} \frac{\mathrm d z}{2\I\pi} \ratioGammaLargeTime(z) e^{-xz+ \frac{z^3}{3}},
\label{eq:defABrownian}
\end{equation} 
	where the contour $\Gamma_A$ is a vertical line with real part between $0$ and $\min \lbrace \Arescaled, \Brescaled , \Crescaled \rbrace $. Note that the function  $A^{(\Arescaled, \Brescaled, \Crescaled)}$ has exponential decay at $+\infty$, that is for any $d\in (0, \min \lbrace \Arescaled, \Brescaled , \Crescaled\rbrace)$, there exist $C\in \R$ such that $\left\vert A^{(\Arescaled, \Brescaled, \Crescaled)}(x)\right\vert\leqslant Ce^{-dx}$. 	Let us introduce an operator $\hat A_s$ acting on $\mathbb L^2(0,+\infty)$ with kernel
\begin{equation} \label{khatA} 
A_s(x,y)= A^{(\Arescaled, \Brescaled, \Crescaled)}(x+y+s), 
\end{equation}
 and an operator $\bar{K}^{(a,b, \Crescaled)}_s$  acting on $\mathbb L^2(0,+\infty)$ with kernel 
$\bar{K}^{(a,b,c)}_s(x, y) := \bar{K}^{(a,b,c)}(x+s, y+s)$. For any $s\in \R$, and $\Arescaled, \Brescaled, \Crescaled>0$, we have \cite[Claim 7.1]{barraquand2020half}
\begin{equation}
\sqrt{\det(I - \bar K^{(\Arescaled, \Brescaled, \Crescaled)}_s)} = \frac{1}{2} \left(\det(I - \hat A_s) + \det(I + \hat A_s)\right),
\end{equation} 
where all operators act on $\mathbb L^2(0,+\infty)$. Hence one has 
\be
G^{(\Arescaled, \Brescaled, \Crescaled)}(s) = \frac{1}{2} \left(1 + \frac{\partial_s}{\Brescaled+\Crescaled}\right)   \left(\det(I - \hat A_s) + \det(I + \hat  A_s)\right)
\label{eq:defGabc}
\ee 
In order to compute $G^{(\Arescaled, \Brescaled, \Crescaled)}(s)$, we had assumed that $a,b,c>0$. However, if we fix some $a>0$, we expect that the RHS of \eqref{eq:limproba} is analytic in $b$ and $c$ in the region $\lbrace (b,c)\in \mathbb R^2; b+c>0, a+b>0, a+c>0\rbrace$ (recall that when $a+b\leq 0$ or  $a+c\leq 0$, the initial condition would not even be defined, and when $b+c\leq 0$ we cannot compute the moments). In the sequel, we implicitly extend the definition of $G^{(\Arescaled, \Brescaled, \Crescaled)}(s)$ by analyticity. To be more precise, $G^{(a,b,c)}(s)$  is defined in \eqref{eq:defGabc} in terms of the operator $\hat A_s$ with  kernel $A^{(\Arescaled, \Brescaled, \Crescaled)}(x+y+s)$ where the function $A^{(\Arescaled, \Brescaled, \Crescaled)}$ is defined in \eqref{eq:defABrownian}. This function can be readily extended analytically to $b<0$ (for example). Indeed, the value when $b<0$ can be expressed by first moving the integration contour to the right of $b$, taking into account the associated residue, and finally setting $b$ to the desired negative value. 

\subsection{Stationary limit $c\to -b$} 
\label{sec:stationarylimit} Assume that $a>b,c,0$. 
In this section we set $c=-b+\epsilon$ and let $\epsilon$ go to $0$ to obtain the stationary limit.  Let us rewrite 
\begin{equation}
A^{(\Arescaled, \Brescaled, \Crescaled)}(x) = \tilde A^{(\Arescaled, \Brescaled, \Crescaled)}(x) + 2\frac{b+c}{b-c} \left( h_c(x) - h_b(x) \right), 
\label{eq:defAtildedecomposition}
\end{equation} 
where 
\begin{equation}
    \tilde A^{(\Arescaled, \Brescaled, \Crescaled)}(x)  = \int_{\Gamma_{\tilde A}} \frac{\mathrm d z}{2\I\pi} \ratioGammaLargeTime(z) e^{-xz+ \frac{z^3}{3}},
\end{equation}
the contour being now a vertical line with real part between $\max\lbrace 0,b,c\rbrace $ and $a$, and 
\begin{equation}
    h_b(x) = b\frac{a+b}{a-b} e^{\frac{b^3}{3}-bx}.
\end{equation}
From now on, we use quantum mechanical notations with kets and bra. 
For functions $f,g\in \mathbb L^2(0, +\infty)$, and an operator $O$ on $\mathbb L^2(0, +\infty)$ acting with a kernel $O(x,y)$, we denote by $\bra{f} O \ket{g}$ the integral $\int_0^{+\infty}dx \int_0^{+\infty} dy f(x)O(x,y)g(y)$ and we denote by $\ket{f}\bra{g}$ the operator acting on $\mathbb L^2(\mathbb R_{+})$ with kernel $f(x)g(y)$. In particular, $\bra{1}, \ket{1}$ below corresponds to the constant functions $f(x)=1, g(y)=1$. 

We rewrite the kernel as
\begin{equation}
    A_s(x,y) = \tilde A^{(\Arescaled, \Brescaled, \Crescaled)}_s(x+y) + 2\frac{b+c}{b-c} \left( \ket{ f_c(x)} \bra{ g_c(y)} -\ket{ f_b(x)} \bra{ g_b(y)}\right), 
\end{equation}
where 
\begin{align}
    f_{\alpha}(x)= \alpha\frac{a+\alpha}{a-\alpha}e^{\frac{\alpha^3}{3}-(x+s)\alpha}, \qquad      g_{\beta}(x)= e^{-x \beta}. 
\end{align}
Here $\tilde A^{(\Arescaled, \Brescaled, \Crescaled)}_s(x) = \tilde A^{(\Arescaled, \Brescaled, \Crescaled)}(x+s)$, and we will also denote by the same symbol $\tilde A^{(\Arescaled, \Brescaled, \Crescaled)}_s$ the operator with kernel $\tilde A^{(\Arescaled, \Brescaled, \Crescaled)}_s(x+y)$. 
Then we have 
\begin{multline} 
    \det\left(I \pm \hat A^{(\Arescaled,\Brescaled,\Crescaled)}_s\right) = \\ \det\left( I\pm \tilde A^{(\Arescaled, \Brescaled, \Crescaled)}\right) \left( \left(1\pm 2\frac{b+c}{b-c}I_{cc} \right) \left(1\mp 2\frac{b+c}{b-c}I_{bb} \right) +4\left( \frac{b+c}{b-c}\right)^2 I_{bc}I_{cb}\right)
\end{multline}
where 
\begin{equation}
    I_{\alpha, \beta}  = \braket{ f_{\alpha} \vert g_{\beta} } \mp \bra{f_{\alpha} } \frac{\tilde A^{(\Arescaled, \Brescaled, \Crescaled)}_s}{I\pm \tilde A^{(\Arescaled, \Brescaled, \Crescaled)}_s} \ket{ g_{\beta}}
\end{equation}
Explicitly, we have 
\begin{equation}
    I_{\alpha, \beta}= e^{\frac{\alpha^3}{3} -s\alpha} \frac{a+\alpha}{a-\alpha}\left( \frac{\alpha}{\alpha+\beta} \mp   \alpha R_{a,b,c}^{\pm}(\alpha, \beta) \right),
\end{equation}
where 
\begin{equation} R_{a,b,c}^{\pm}(\alpha, \beta) = \bra{ e^{-x\alpha}} \frac{\tilde A^{(\Arescaled, \Brescaled, \Crescaled)}_s}{1\pm \tilde A^{(\Arescaled, \Brescaled, \Crescaled)}_s} \ket{e^{-x\beta}}. 
\end{equation}

where the braket notation denotes the two-sided Laplace transform and has the following definition: for any operator $\mathcal{O}$ acting on $\mathbb{L}^2(\R_+)$ with kernel $(u,v)\mapsto \mathcal{O}(u,v)$ we have
\begin{equation}
    \bra{ e^{-x\alpha} } \mathcal{O} \ket{ e^{-x\beta} }=\iint_{\R_+^2} \rmd u \rmd v \, e^{-\alpha u}\mathcal{O}(u,v)e^{-\beta v}
\end{equation}

In the limit when $c\to -b$, we find, after simplifications using Mathematica, that 
\begin{equation}\label{eq:Fab}
   \lim_{b\to -c}F^{(a,b,c)}(s) = G_{a,b}^{\rm HY}(s) := \frac{1}{2} \partial_s \left( \det(1+\tilde A_s) Q^+(a,b,s) +  \det(1-\tilde A_s) Q^-(a,b,s)\right),
\end{equation}
where $\tilde A_s$ is now an operator acting with kernel 
$ \tilde A_s(x,y) =\tilde A(s+x+y)$ with 
\begin{equation}
\tilde A(x)  =  \int \frac{\mathrm d z}{2\I\pi} \frac{\Arescaled +z}{\Arescaled -z}   e^{-xz+ \frac{z^3}{3}},
\label{eq:defAtilde}
\end{equation} 
where the contour is a vertical line with real part between $0$ and $a$. We define 
$$Q^{\pm}(a,b,s)= S^{\pm}(a,b,s)+S^{\pm}(a,-b,s),$$ 
where
\begin{align} 
    S^+(a,b,s)  &= \frac{1}{2} \left(s-b^2+2R^+(b,-b)  \right) +\frac{(a+b)^2}{2b(a^2-b^2)}  e^{b^3/3-bs} (2bR^+(b,b)-1) -\frac{a}{a^2-b^2} , \\
    S^-(a,b,s)  &= \frac{1}{2} \left( s-b^2-2R^-(b,-b) \right) +\frac{(a+b)^2}{2b(a^2-b^2)} e^{b^3/3-bs}(2bR^-(b,b)+1) -\frac{a}{a^2-b^2}. 
\end{align}
and 
\begin{equation} R^{\pm}(\alpha, \beta) = \bra{ e^{-x\alpha}}  \frac{\tilde A_s}{1\pm \tilde A_s} \ket{ e^{-x\beta} }. 
\end{equation}

\subsection{Case $b=0, a>0$}
This case corresponds to the maximal current phase. Only the $+$ term remains, as in \cite{barraquand2020half}. 
In the special case $b=0$, we have the simplifications
\begin{align}
    Q^+(a,0,s)&= \frac{2}{a} \left(-2+as +2a R^+(0,0) \right),\\ 
    Q^-(a,0,s)&= 0,
\end{align}
so that 
\begin{equation}\label{eq:defFa0}
G_{a,0}^{\rm HY}(s) =\partial_s \left( \det(I+\tilde A_s)\left( -\frac{2}{a}+s +2R^+(0,0) \right)\right). 
\end{equation}

\begin{remark}
\label{rem:idenityindistribution}
As $a\to +\infty$, we recover $F^{\rm Brownian}_a$ studied in \cite{barraquand2020half} (denoted simply $F$ in \cite{barraquand2020half}). This is due to a  symmetry between boundary and initial condition parameters \cite[Section 4.6]{barraquand2020half} (see also  \cite{parekh2017kpz, barraquand2020halfMacdonald}). The law of $h(0,t)$ for the half-line KPZ equation with Dirichlet boundary condition (that is $a=+\infty$) and with initial condition given by a O'Connell-Yor polymer partition function,  is the same as the law of $h(0,t)$ for the half-line KPZ equation with Robin boundary condition and  Brownian initial condition, for appropriately chosen parameters, see details in  \cite[Sections 4.5 and 4.6]{barraquand2020half}. More generally, using the results of \cite[Sections 4.6]{barraquand2020half} we obtain  that $\lim_{a\to\infty}G_{a,b}^{\rm HY} =  F^{\rm Brownian}_{-b}$.
\end{remark} 

\section{Two-time covariance} 
\label{sec:twotime}
\subsection{Full-space variational formulas} 
In full space one has, for the droplet initial condition $e^{h(x,0)}=\delta_0(x)$,  at large time $t$, 
\be 
h(x,t)+\frac{t}{12} \simeq t^{1/3} ({\cal A}_2(\hat x) - \hat x^2), \quad  \quad \hat x = \frac{x}{2 t^{2/3}} 
\ee 
For a Brownian IC one has 
\be 
h(x,t)+\frac{t}{12} \simeq t^{1/3} \mathcal A^{\rm stat}(\hat x) 
\label{eq:Airystat}
\ee 
where the process $\mathcal A^{\rm stat}$ was introduced in \cite{baik2010limit}, and can be characterized by the following formula \cite{quastel2014airy}:  for any fixed $\hat x$, 
\be 
\mathcal A^{\rm stat}(\hat x) = \max_{\hat y\in \mathbb R} (\sqrt{2} B(\hat y) + {\cal A}_2(\hat x- \hat y) - (\hat x -\hat y)^2 ).
\ee 
Since the Brownian motion is stationary for the KPZ equation, \eqref{eq:Airystat} implies that, as processes in $\hat x$, 
\begin{equation}
    \mathcal A^{\rm stat}(\hat x)-\mathcal A^{\rm stat}(0) \overset{(d)}{=}  \sqrt{2} B(\hat x).
\end{equation}
where $B$ is a two-sided Brownian motion.

\subsection{Half-space universal processes}

In half-space, for droplet IC and boundary parameter $u=at^{-1/3}$ one has, for large $t$ and fixed $a$, the solution $h_u(x,t)$ behaves as 
\begin{equation}  \label{eq:hudrop} 
h_{u=a t^{-1/3}}(x,t)+\frac{t}{12} \simeq t^{1/3} ({\cal A}_a(\hat x) - \hat x^2), \quad \quad \hat x = \frac{x}{2 t^{2/3}}\geqslant 0,
\end{equation} 
where ${\cal A}_a(\hat x)$ is a half-space variant of the Airy$_2$ process, having explicit finite-dimensional marginal distributions computed in \cite{baik2018pfaffian} (this limiting process was obtained as a limit of a model of last passage percolation in a half-space, but by universality, the same should arise as a limit of the KPZ equation).  More generally, for an  initial condition $h_0(x)=h(x,0)$, such that the rescaled process  
\begin{equation}
    h^{\rm rescaled}_0(x)  = \lim_{y\to\infty} \frac{1}{\sqrt{y}} h_0(2xy)
\end{equation}
exists, we expect that the solution $h_u(x,t)$ of the half-space KPZ equation with boundary parameter $u$ behaves asymptotically as 
\begin{equation} \label{eq:evolutiona} 
h_{u=a t^{-1/3}}(0,t)+\frac{t}{12} \simeq t^{1/3} \max_{\hat y \geqslant 0} \left\lbrace h_0^{\rm rescaled}(\hat y) + {\cal A}_a(\hat y) - \hat y^2 \right\rbrace.
\end{equation}

For the stationary initial condition $h(x,0)= h^{\rm stat}_{u,v}(x)$, we similarly expect that 
for $u=a t^{-1/3}$ and $v=b t^{-1/3}$, there exists a process $\mathcal A^{\rm stat}_{a,b}(\hat x)$ such that at large $t$
\begin{equation}  \label{eq:hA} 
h_{at^{-1/3}}(x,t) +\frac{t}{12}\simeq t^{1/3} \mathcal A^{\rm stat}_{a,b}(\hat x), \quad \quad \hat x = \frac{x}{2 t^{2/3}}\geqslant 0.
\end{equation} 
In particular, we have that 
\begin{equation}  \label{eq:Axi} 
\mathcal A^{\rm stat}_{a,b}(0)\overset{(d)}{=} \xi_{a,b}.
\end{equation} 
In the case $a+b=0$ and $a\leq 0, a\leq b$, that is when the initial condition is Brownian this process is defined and studied in \cite{betea2020half}. 
We expect that 
\begin{equation}
    \mathcal A^{\rm stat}_{a,b}(0) = \max_{\hat y\geq 0} \left\lbrace  h^{\rm rescaled}_{a,b}(\hat y)  + {\cal A}_a(\hat y) - \hat y^2 \right\rbrace.
    \label{eq:xivariational}
\end{equation}
where the process $h^{\rm rescaled}_{a,b}$ is defined as 
\be
h^{\rm rescaled}_{a,b}(x) = \lim_{r\to\infty} \frac{1}{\sqrt{r}}h^{\rm stat}_{ar^{-1/2},br^{-1/2}}(2r x).
\ee
We may describe this process very explicitly using \eqref{eq:defhstat}, even in cases where it is not Brownian.  For $a\geq b, b\leq 0$, we have that 
\begin{equation}
    h^{\rm rescaled}_{a,b}(x) = \max\left\lbrace   B_2(2x) , -E_{a-b}+\max_{t\in [0,x]}\left\lbrace B_1(2t)+B_2(2x)-B_2(2t)\right\rbrace \right\rbrace , 
\end{equation}
where $B_1, B_2$ are independent standard Brownian motions with drifts $-b$ and $b$ respectively, and $E_{a-b}$ is an independent exponential random variables with parameter $a-b$. For $a\geqslant 0, b\geq 0$, 
\begin{equation}
    h^{\rm rescaled}_{a,b}(x) = \max\left\lbrace   B_2(2x) , -E_{a}+\max_{t\in [0,x]}\left\lbrace B_1(2t)+B_2(2x)-B_2(2t)\right\rbrace \right\rbrace, 
\end{equation}
where $B_1, B_2$ are independent standard Brownian motions (without drift). In the limit $a \to +\infty$ and $b=0$,  this process  has the same law as the maximum of two non intersecting Brownians \cite{o2002representation}.
 For $a\leq b, a\leq 0$, 
\begin{equation}
    h^{\rm rescaled}_{a,b}(x) = B(x)+ax, 
\end{equation}
where $B$ is a standard Brownian motion. \\

By stationarity, we have the equality in distribution of processes in the variable $x$: 
\begin{equation}
\mathcal A^{\rm stat}_{a,b}(x)- \mathcal A^{\rm stat}_{a,b}(0) \overset{(d)}{=} h^{\rm rescaled}_{a,b}(x). 
\label{eq:usingstationarity}
\end{equation}

\subsection{Computation of two-time covariances}
In this Section we compute two-time covariances starting from various initial conditions. 
To this purpose we adapt the argument from \cite{ferrari2016time} to the half-space geometry. 
We also use similar notations as in that paper.
Let us use the notation, where $t$ is the late time and $\tau t$ the earlier time, $0< \tau < 1$, 
\begin{equation}
    C(\tau ) = \lim_{t\to\infty}\mathrm{Cov}\left(\mathcal{X}_t(\tau), \mathcal{X}_t(1)  \right), \,\,\, \mathcal{X}_t(\tau) = \frac{h_u(0,t\tau)+\frac{t\tau }{12}}{t^{1/3}}. 
\end{equation}
We will use the formula 
\begin{equation}\label{eq:generalcovariance}
\mathrm{Cov}\left(\mathcal{X}_t(\tau), \mathcal{X}_t(1)  \right)= \frac{1}{2}\Var\left[ \mathcal X_t(1) \right] + \frac{1}{2}\Var\left[ \mathcal X_t(\tau) \right] - \frac{1}{2}\Var\left[ \mathcal X_t(1)-X_t(\tau)  \right].
\end{equation}

\subsubsection{Stationary Hariya-Yor initial condition}
Here we relate the two-time covariance of the scaled KPZ height field with stationary initial condition to the variance of the random variable $\xi_{a,b}$
studied in this paper of CDF denoted $F^{\rm stat}_{a,b}$, defined in \eqref{eq:defFabstat}. 

For the moment we focus on the regions in the regions $R_1$ and $R_3$, that is $u> v, v\leq 0$.  Assume that we start from  the initial condition $h(x,0)=h^{\rm stat }_{u,v}(x)$. 
Recall that by stationarity, $\mathbb E \left[\mathcal{X}_t(\tau)\right]=0$, so that we have
\begin{equation} \label{eq:Ctau2} 
    C(\tau) = \lim_{t\to\infty}\frac{1}{2} \left( \Var \mathcal{X}_t(1) + \Var \mathcal{X}_t(\tau) -\mathbb E\left[ (\mathcal{X}_t(1)-\mathcal{X}_t(\tau))^2 \right]  \right) .
\end{equation}
Scaling $u,v$ as  $u=a t^{-1/3}$, $v=bt^{-1/3}$, we have, by replacing $t \to \tau t$ and $a \to a \tau^{1/3}$, $b \to b \tau^{1/3}$ in \eqref{eq:hA} 
that the height field at the earlier time satisfies
\be \label{eq:ha} 
h_{at^{-1/3}}(x, \tau t) + \frac{\tau t}{12} \simeq  (\tau t)^{1/3} \mathcal A^{\rm stat}_{a \tau^{1/3},b \tau^{1/3}}\left(\frac{x}{2 (\tau t)^{2/3}}\right) 
\ee 
This field can then be used as an initial condition for the evolution from time $\tau t$ to time $t$. Using formula \eqref{eq:evolutiona} with $t \to (1-\tau) t$, 
$a \to a (1-\tau)^{1/3}$ and $\hat y= \frac{x}{2 (t (1-\tau))^{2/3}} $
it leads to the variational formula, where we denote $\hat \tau=\frac{\tau}{1-\tau}$
\begin{equation}
    \mathcal{X}_t(1) \simeq (1-\tau)^{1/3} \max_{\hat y>0} \left\lbrace \hat \tau^{1/3} \mathcal A_{a\tau^{1/3},b\tau^{1/3}}^{\rm stat}\left( \hat\tau^{-2/3} \hat y\right) + \mathcal A_{a(1-\tau)^{1/3}}(\hat y)- \hat y^2\right\rbrace, 
\end{equation}
where the processes $\mathcal A_{a\tau^{1/3},b\tau^{1/3}}^{\rm stat}$ and $\mathcal A_{a(1-\tau)^{1/3}}$ are independent, because  they describe the growth over two disjoint time intervals.  
Thus, using \eqref{eq:ha} with $x=0$, $\mathcal{X}_t(1) -\mathcal{X}_t(\tau)$ has asymptotically the same distribution as 
\begin{equation}\label{eq:diffvariational}
    (1-\tau)^{1/3} \max_{\hat y>0} \left\lbrace \hat \tau^{1/3} \left(\mathcal A_{a\tau^{1/3},b\tau^{1/3}}^{\rm stat}\left( \hat\tau^{-2/3}\hat y\right)-\mathcal A_{a\tau^{1/3},b\tau^{1/3}}^{\rm stat}\left( 0\right)\right) + \mathcal A_{a(1-\tau)^{1/3}}(\hat y)-\hat y^2\right\rbrace.
\end{equation}
Using \eqref{eq:usingstationarity} and \eqref{eq:xivariational}, we obtain that \eqref{eq:diffvariational} has the same distribution as 
\begin{equation}
  (1-\tau)^{1/3} \max_{\hat y>0} \left\lbrace \hat \tau^{1/3} h^{\rm rescaled}_{a\tau^{1/3},b\tau^{1/3}}( \hat\tau^{-2/3}\hat y) + \mathcal A_{a(1-\tau)^{1/3}}(\hat y)-\hat y^2\right\rbrace.
  \end{equation}
Now, we use the fact that the process $h^{\rm rescaled}$ satisfies the scaling property 
\be \label{eq:rescHY} 
\frac{1}{\sqrt{r}}h^{\rm rescaled}_{ar^{-1/2},br^{-1/2}}(rx)\overset{(d)}{=}h^{\rm rescaled}_{a,b}(x),
\ee
so that using \eqref{eq:evolutiona}, 
\begin{align*} 
\mathcal{X}_t(1) -\mathcal{X}_t(\tau)  &\overset{(d)}{=} (1-\tau)^{1/3} \max_{x>0} \left\lbrace  h^{\rm rescaled}_{a(1-\tau)^{1/3},b(1-\tau)^{1/3}}( x) + \mathcal A_{a(1-\tau)^{1/3}}(x)-x^2\right\rbrace,\\
    &\overset{(d)}{=}  (1-\tau)^{1/3}  \xi_{a(1-\tau)^{1/3},b(1-\tau)^{1/3}}.
\end{align*}
where in the last formula we used \eqref{eq:xivariational} and \eqref{eq:Axi}.
Finally, we obtain that using \eqref{eq:hA} and \eqref{eq:Ctau2} 
\begin{equation}
C(\tau) = \frac{1}{2} \left( \Var (\xi_{a,b}) + \tau^{2/3} \Var (\xi_{a\tau^{1/3}, b\tau^{1/3}}) - (1-\tau)^{2/3}\Var (\xi_{a(1-\tau)^{1/3},b(1-\tau)^{1/3}})\right).
\end{equation}

\subsubsection{Droplet initial condition}

Now we consider the solution $h_u$ of the half-space KPZ equation with boundary parameter $u$ and droplet initial condition at the origin. At large time, the height field should converge locally to one of the invariant distributions. The limiting distribution depends on the boundary parameter, $u$, and the drift of the initial condition, according to the diagram in \cite[Fig.~2]{barraquand2021kardar}. In the case of the droplet initial condition, the  drift parameter $v=+\infty$, so the height field converges to the invariant process $\mathcal{HY}_{u,0}$, see \cite[Fig.~2]{barraquand2021steady}. More precisely, for $x$ in a domain of order $1$, and $u>0$, 
\begin{equation}
    \lim_{t\to \infty} h_u(x,t)-h_u(0,t) \overset{(d)}{=} \mathcal{HY}_{u,0}(x)-\mathcal{HY}_{u,0}(0).
    \label{eq:locallimit1}
\end{equation}
When $u<0$, 
\begin{equation}
    \lim_{t\to \infty} h_u(x,t)-h_u(0,t) \overset{(d)}{=}B(x)+ux.
    \label{eq:locallimit2}
\end{equation}
Hence, for any $u$, 
\begin{equation}
    \lim_{t\to \infty} h_u(x,t)-h_u(0,t) \overset{(d)}{=} h^{\rm stat}_{u,0}(x).
    \label{eq:locallimit}
\end{equation}

On the other hand, we also know by \eqref{eq:hudrop} that on the scale $t^{2/3}$, if $u=at^{-1/3}$ the height field converges to $\mathcal A_a$, that is 
\begin{equation}
    \lim_{t\to \infty} h_u(2\hat xt^{2/3},t)-h_u(0,t) \overset{(d)}{=} t^{1/3} \left(\mathcal{A}_{a}(\hat x)-\mathcal{A}_{a}(0)-\hat x^2\right).
    \label{eq:steprescaled}
\end{equation}
We expect that \eqref{eq:locallimit} and \eqref{eq:steprescaled} match when $x=2\hat xt^{2/3}$ goes to infinity and $\hat x$ goes to zero. This implies that, for $\hat x$ going to zero, 
\begin{equation}
    \lim_{t\to \infty} t^{-1/3} h^{\rm stat}_{at^{-1/3}}(2\hat xt^{2/3})  \overset{(d)}{=}  \mathcal A_a(\hat x)-\mathcal A_a(\hat x). 
\end{equation}
In other terms, for $\hat x$ going to zero and any fixed $a\in \mathbb R$, 
\begin{equation} \label{eq:AaHY} 
    h^{\rm rescaled}_{a,0}(\hat x) \simeq   \mathcal A_a(\hat x)-\mathcal A_a(0).
\end{equation}

Scaling $u$ as  $u=a t^{-1/3}$, we obtain, by replacing $t \to \tau t$ and $a \to a \tau^{1/3}$ in \eqref{eq:hudrop}, 
that the height field at the earlier time satisfies for large $t$
\be \label{eq:habis} 
h_{at^{-1/3}}(x, \tau t) + \frac{\tau t}{12} \simeq  (\tau t)^{1/3} \left(  \mathcal A_{a \tau^{1/3}} \left(\frac{x}{2 (\tau t)^{2/3}}\right) - \frac{x^2}{4 (\tau t)^{4/3}} \right)
\ee 
This field can then be used as an initial condition for the evolution from time $\tau t$ to time $t$. Using formula \eqref{eq:evolutiona} with $t \to (1-\tau) t$, 
$a \to a (1-\tau)^{1/3}$ and $\hat y= \frac{x}{2 (t (1-\tau))^{2/3}} $
it leads to the variational formula, where we denote $\hat \tau=\frac{\tau}{1-\tau}$
\begin{equation}
    \mathcal{X}_t(1) \simeq (1-\tau)^{1/3} \max_{\hat y>0} \left\lbrace \hat \tau^{1/3} \mathcal A_{a\tau^{1/3}}\left( \hat\tau^{-2/3} \hat y\right)
    - \hat \tau^{-1} \hat y^2 + \tilde{\mathcal A}_{a(1-\tau)^{1/3}}(\hat y)- \hat y^2\right\rbrace, 
\end{equation}
where the processes $\mathcal A_{a\tau^{1/3}}$ and $\tilde{\mathcal A}_{a(1-\tau)^{1/3}}$ are independent, because  they describe the growth over two disjoint time intervals.  
Thus, using \eqref{eq:habis} with $x=0$, $\mathcal{X}_t(1) -\mathcal{X}_t(\tau)$ has asymptotically the same distribution as 
\begin{equation}\label{eq:diffvariationalbis}
    (1-\tau)^{1/3} \max_{\hat y>0} \left\lbrace \hat \tau^{1/3} \left(\mathcal A_{a\tau^{1/3}}\left( \hat\tau^{-2/3}\hat y\right)-\mathcal A_{a\tau^{1/3}} \left( 0\right) - \hat\tau^{-4/3}\hat y^2 \right) + \tilde{\mathcal A}_{a(1-\tau)^{1/3}}(\hat y)-\hat y^2\right\rbrace.
\end{equation}
In the limit $1-\tau \ll 1$, the argument $\hat\tau^{-2/3}\hat y$ is small and one can use \eqref{eq:AaHY} and the scale invariance property
\eqref{eq:rescHY} leading to 
\begin{align*} 
    \mathcal{X}_t(1)-\mathcal X_t(\tau) &\simeq (1-\tau)^{1/3} \max_{\hat y>0} \left\lbrace h^{\rm rescaled}_{a(1-\tau)^{1/3},0}(\hat y) + \tilde{\mathcal A}_{a(1-\tau)^{1/3}}(\hat y)- \hat y^2\right\rbrace, \\
    &\simeq  (1-\tau)^{1/3}  \xi_{a(1-\tau)^{1/3},0}
\end{align*}
where in the last formula we used \eqref{eq:xivariational} and \eqref{eq:Axi} . 
So that, the formula \eqref{eq:generalcovariance} yields, for any  $a\in \mathbb R$ and $\tau\to 1$, 
\begin{equation}
    C(\tau) = \frac{1}{2}\Var\left[ \mathcal A_a(0) \right] + \frac{1}{2}\tau^{2/3}\Var\left[ \mathcal A_{a\tau^{1/3}}(0) \right] - \frac{1}{2}(1-\tau)^{2/3}  \Var\left[\xi_{a(1-\tau)^{1/3},0}\right] + \mathcal O(1-\tau).
\end{equation}
Since this formula is valid only in the limit $\tau\to 1$, we may simplify it using $\Var[\xi_{a(1-\tau)^{1/3},0}] = \Var[\xi_{0,0}]+\mathcal O((1-\tau)^{1/3})$ and $\Var\left[ \mathcal A_{a\tau^{1/3}}(0) \right]=\Var\left[ \mathcal A_{a}(0) \right]+ \mathcal O(1-\tau)$, so that 
\begin{equation}
    C(\tau) = \Var\left[ \mathcal A_a(0) \right] - \frac{1}{2}(1-\tau)^{2/3}  \Var\left[\xi_{0,0}\right] + \mathcal O(1-\tau),
\end{equation}
thus we obtain \eqref{eq:introcorrelationdroplet} as announced. 

\newpage

\appendix 

\begin{center}
    \Large \bf
    Appendix
\end{center}

\section{Brownian case}
\label{sec:Brownian}
In \cite{barraquand2020half}, we computed only the function $F_a^{\rm Brownian}(s)$ when $a=0$. We show in this Section that very similar arguments as those already developed in \cite{barraquand2020half} also yields an expression for $F_a^{\rm Brownian}(s)$ for any $a\in \mathbb R$. We start from 
\begin{equation}
    F_a^{\rm Brownian}(s)= \lim_{b\to -a} F^{(\Arescaled, \Brescaled)}(s)
    \label{eq:defFabtoaBrownian}
\end{equation}
where $F^{(\Arescaled, \Brescaled)}(s)$ was given in \cite[(7.11)]{barraquand2020half} as
\be
\label{eq:Fabnosquareroot}
F^{(\Arescaled, \Brescaled)}(s) = \frac{1}{2} \left(1 + \frac{\partial_s}{\Arescaled+\Brescaled}\right)   \left(\det(I - \hat A_s) + \det(I + \hat A_s)\right)
\ee
We will follow the same notations as in \cite{barraquand2020half} (up to minor changes). The operator $A_s$ acts $\mathbb L^2(0,+\infty)$ with kernel
\begin{equation} 
\hat A_s(x,y)= A^{(\Arescaled, \Brescaled)}(x+y+s), 
\end{equation}
where the function $A^{(\Arescaled, \Brescaled)}(x)$ is defined by 
\begin{equation}
A^{(\Arescaled, \Brescaled)}(x) = \int \frac{\mathrm d z}{2\I\pi} \frac{\Arescaled +z}{\Arescaled -z}  \frac{\Brescaled+z}{\Brescaled -z} e^{-xz+ \frac{z^3}{3}},
\label{eq:defAexplicit}
\end{equation} 
and the contour is a vertical line with real part between $0$ and $\min \lbrace \Arescaled, \Brescaled \rbrace $. Moving the contour to the right, we obtained in \cite{barraquand2020half} that 
\begin{equation} 
A^{(\Arescaled, \Brescaled)}(x) = \tilde A^{(\Arescaled, \Brescaled)}(x) + 2 \frac{\Arescaled+\Brescaled}{\Arescaled-\Brescaled} (h_\Brescaled(x) - h_\Arescaled(x) ),
\end{equation}
with  $h_\Brescaled(x)=\Brescaled e^{- x \Brescaled + \Brescaled^3/3}$. Letting $b=-a+\epsilon$, we have 
\begin{equation}
\tilde A^{(\Arescaled, \Brescaled)}(x) = \int \frac{\rmd z}{2 \I \pi} \frac{\Arescaled+z}{\Arescaled-z} \frac{\Brescaled+z}{\Brescaled-z} e^{-x z + z^3/3} 
= {\rm Ai}(x) + 2 \epsilon \int_x^{+\infty}\rmd \lambda \, \cosh(ay)\Ai(\lambda+y) + \mathcal{O}(\epsilon^2), 
\end{equation}
\note{I corrected that formula} 
where in the integral over $z$, the contour passes to the right of $a, b$. We also introduce the operator $\tilde A_s$ acting on $\mathbb L^2(0,+\infty)$ with kernel 
$$ \tilde A_s (x,y)= \tilde A^{(\Arescaled, \Brescaled)}(s+x+y).$$
$\hat A_s$ is a rank-$2$ perturbation of  $\tilde A_s$, in the sense that 
\be
\hat A_s(x,y)= \tilde A_s(x,y) + 2 \frac{\Arescaled+\Brescaled}{\Arescaled-\Brescaled} ( b\ket{f_\Brescaled(x) } \bra{ f_\Brescaled(y) } 
-  a\ket{f_\Arescaled(x) } \bra{ f_\Arescaled(y) } ), 
\ee 
with $f_\Brescaled(x) =  e^{\Brescaled^3/6 - \Brescaled s/2 -\Brescaled x}$. Using the matrix determinant lemma, we have 
\begin{equation} \label{dets} 
\det(I  \mp  \hat A_s) = 
\det(I \mp \tilde A_s) \left( \left(1 \mp 2 b\frac{a+b}{a-b} I_{b,b} \right) 
\left(1 \pm 2 a\frac{a+b}{a-b} I_{a,a} \right) + 4ab \left(\frac{a+b}{a-b}\right)^2
I_{b,a} I_{a,b} \right) 
\end{equation} 
where 
\be
I_{\alpha,\beta} = \braket{ f_\alpha | f_\beta } \pm \bra{ f_\alpha } \frac{\tilde A_s}{I \mp \tilde A_s}  \ket{ f_\beta }.
\ee 

The scalar products are evaluated as 
\be
\braket{ f_\alpha | f_\beta } = \frac{1}{\alpha + \beta}  e^{\frac{\alpha^3}{6} + \frac{\beta^3}{6} - \frac{\alpha+\beta}{2} s} 
\ee 
and 
\be
\bra{f_\alpha }\frac{\tilde A_s}{I \mp \tilde A_s} \ket{ f_\beta }
= e^{\frac{\alpha^3}{6} + \frac{\beta^3}{6} }
e^{- \frac{\alpha+\beta}{2} s} 
\bra{e^{-\alpha x} } \frac{\tilde A_s}{I \mp \tilde A_s}  \ket{ e^{-\beta x}}.
\ee 
Putting all this into Mathematica, we find that 
\be \det(I - \hat A_s) + \det(I + \hat A_s)\ee
is of order $\epsilon$. Hence, dividing by $\epsilon$ and  using \eqref{eq:defFabtoaBrownian} and \eqref{eq:Fabnosquareroot}, 
\begin{equation}
    F_a^{\rm Brownian}(s)= \frac{1}{2} \partial_s \left( \det(I-\Ai_s) S^-_a + \det(I+\Ai_s) S^+_a \right) 
    \label{eq:formulaforFaBrownian}
\end{equation}
where $\Ai_s$ denotes the operator with kernel $\Ai(x+y+s)$,  
\be 
S^-_a = e^{a s-\frac{a^3}{3}}
   R^-_{-a,-a}+e^{\frac{a^3}{3}-a s}
   R^-_{a,a}-2 R^-_{a,-a}+\frac{\sinh
   \left(\frac{1}{3} a \left(a^2-3
   s\right)\right)}{a}-a^2+s, 
\ee
\be 
S^+_a = e^{a s-\frac{a^3}{3}}
   R^+_{-a,-a}+e^{\frac{a^3}{3}-a s}
   R^+_{a,a}+2 R^+_{a,-a}-\frac{\sinh
   \left(\frac{1}{3} a \left(a^2-3
   s\right)\right)}{a}-a^2+s
\ee
and 
\be R^{\mp}_{\alpha, \beta} = \bra{e^{-\alpha x} } \frac{ \Ai_s}{I \mp  \Ai_s}  \ket{ e^{-\beta x}}. 
\ee
We may check that for $a=0$, $S^-_0=0$ and $S^+_0=4R^+_0+2s$, so that we recover exactly the result from \cite{barraquand2020half}. Using the Sherman-Morrison formula, we have that 
\begin{equation}
R^{\mp}_{\alpha, \beta} = \mp  \frac{\det(I \mp \Ai_s \mp \ket{ \Ai_s e^{-\beta y} } \bra{   e^{-\alpha x}})}{\det(I \mp \Ai_s)}\pm 1.
\end{equation}
so that $F_a^{\rm Brownian}$ can be written in terms of Fredholm determinants and simple functions, and could be evaluated numerically (to compute the Fredholm determinants of rank one perturbation of $\Ai_s$, one may need to conjugate the kernel so that all kernels involved are decaying at infinity).

\section{Limiting one-point distribution away from the wall at $x>0$}
\label{sec:xpositif}
In this Appendix we study the distribution of the height $h(x,t)$ at $x>0$. 

\subsection{Moment formula}
We start from the moment formula  \eqref{eq:momentformula}, that is 
\begin{multline}
	\EE[Z(x,t)^k] =2^k \frac{\Gamma(v_1+v_2)}{\Gamma(v_1+v_2-k)} \int_{r_1+\I\R}\frac{\mathrm{d}z_1}{2\I\pi}\cdots \int_{r_k+\I\R}\frac{\mathrm{d}z_k}{2\I\pi} \prod_{1\leqslant a<b\leqslant k} \frac{z_a-z_b}{z_a-z_b-1}
	F(\vec z) \\
F(\vec z) = \prod_{1\leqslant a<b\leqslant k}   \frac{z_a+z_b}{z_a+z_b-1} 
	\prod_{i=1}^k \frac{z_i}{z_i+u-1/2}  \frac{1}{(v_1-1/2)^2-z_i^2}\frac{1}{(v_2-1/2)^2-z_i^2}e^{tz_i^2 -  z_i x},
	\label{eq:momentsKPZhalfspaceBrownian}
\end{multline}

We use \cite[Proposition 5.1]{borodin2016directed} %\cite[Proposition 5.1]{borodin2016directed}
specializing to a function $F(\vec z)$ which is symmetric in its arguments
\begin{multline}
\int_{r_1+\I\R}\frac{\mathrm{d}z_1}{2\I\pi}\cdots \int_{r_k+\I\R}\frac{\mathrm{d}z_k}{2\I\pi} 
\prod_{1\leqslant a<b\leqslant k} \frac{z_a-z_b}{z_a-z_b-1} F(\vec z) \\
 = 
 k!  \sum_{\la\vdash k} \frac{1}{m_1(\la)!m_2(\la)!\dots} \int_{a_w+\I\R}\frac{\rmd w_1}{2\I\pi} 
 \dots \int_{a_w+\I\R}\frac{\rmd w_{\ell(\lambda)}}{2\I\pi}  
\det\left[\frac{1}{w_i+\la_i-w_j}\right]_{i,j=1}^{\ell(\lambda)}  \\
\times 
F(w_1,w_1+1,\dots,w_1+\lambda_1-1,\dots,w_\ell(\lambda),w_{\ell(\lambda)}+1,\dots,w_{\ell(\lambda)}
+\lambda_{\ell(\lambda)}-1)
\end{multline}
For the last equation to be valid,  the function $F$ need to be holomorphic in each variable in the region spanned by the contour deformation, that is,  it needs to be holomorphic in the whole region between $r_1+\I\R$ and $r_k+\I\R$. This will be the case if we choose $u-\frac{1}{2}>r_1$ and $r_k>\max\lbrace 0,-v_1+\frac{1}{2},-v_2+\frac{1}{2}\rbrace$ on the left hand side, so that we may take the contour in the right hand side such that $v_1-k+1>a_w>\max\lbrace 1/2-u, 0\rbrace$ (one can simply take $a_w=r_k$).

\subsection{Laplace transform formula}

\begin{figure}
    \centering
    \begin{tikzpicture}[scale=1.6]
\draw[thick, gray, ->] (-1,0) -- (4,0);
\draw[thick, gray, ->] (0,-2) -- (0,3);
\draw[gray] (0,0) node[anchor=north east] {$0$};
\draw[thick] (1.5,0) -- +(60:3.3);
\draw[thick, -stealth] (1.5,0) -- +(60:1);
\draw[thick] (1.5,0) -- +(-60:2.2);
\draw[] (0.5,0.1) -- (0.5,-0.1) node[anchor=north] {$1/2$};
\draw[] (3.5,-0.1) -- (3.5,0.1) node[anchor=south] {$v_1+1/2$};
\draw[] (-0.5,-0.1) -- (-0.5,0.1) node[anchor=south] {$1/2-u$};
\draw[] (1.5,-0.1) -- (1.5,0.1) node[anchor=south] {$a_s$};
%\draw[] (1,0) node[anchor=south east] {$a_w$};
\draw[thick] (1,-2) -- (1,3);
\draw[thick, -stealth] (1,-2) -- (1,-1);
\fill (1,2) circle(0.07);
\fill (2,2) circle(0.07);
\fill (3,2) circle(0.07);
\fill (1.2,2.3) node{$w$};
\fill (2,2.4) node{$w+1$};
\fill (3.2,2.3) node{$w+2$};
\draw[thick, -stealth] (2.25,2) arc(0:-360:0.25);
%\draw[thick, ->] (2.25,2) arc(0:-360:0.25);
\draw (1.5,-1.8) node {$a_w+\I\mathbb R$};
\draw (3,-1.7) node {$\mathcal C_{a_s}[w]$};
\end{tikzpicture}
    \caption{The contour $\mathcal C_{a_s}[w]$ used in \eqref{eq:MellinBarnesxpositif}.}
    \label{fig:contourpluscircles}
\end{figure}
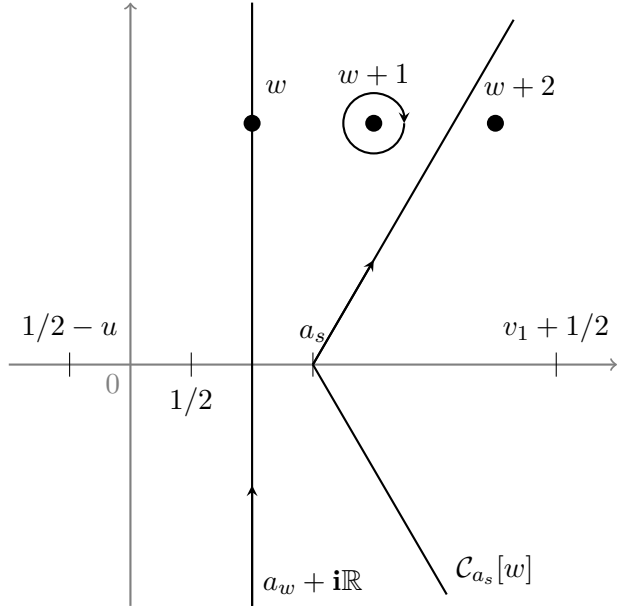

We can evaluate the function $F$ into strings using the same manipulations  as in Section \eqref{sec:Pfaffian}, and using the idenity 
\begin{equation}
\prod_{a=0}^{\lambda_1-1}  \prod_{b=0}^{\lambda_2-1} \frac{w_1+w_2 + a + b}{ w_1+w_2  + a + b -1} = \frac{\Gamma \left(w_1+w_2-1\right) \Gamma \left(w_1+w_2+\lambda _1+\lambda
   _2-1\right)}{\Gamma \left(w_1+w_2+\lambda _1-1\right) \Gamma \left(w_1+w_2+\lambda
   _2-1\right)} 
\end{equation}
and 
\begin{equation} 
 \prod_{0 \leq a < b \leq \lambda-1} \frac{2 w + a + b}{ 2 w + a + b -1} 
=  \frac{2^{-\lambda} \Gamma (w) \Gamma \left(2(w+\lambda) -1 \right)}{\Gamma(w+\lambda) \Gamma (2 w+\lambda -1)}.
\end{equation}
We obtain
\begin{multline}      	\EE[Z(x,t)^k] =\frac{2^k  \Gamma(v_1+v_2)}{\Gamma(v_1+v_2-k)}  
 k!  \sum_{\la\vdash k} \frac{1}{m_1(\la)!m_2(\la)!\dots}  \\ \int_{a_w+\I\R}\frac{\rmd w_1}{2\I\pi} 
 \dots \int_{a_w+\I\R}\frac{\rmd w_{\ell(\lambda)}}{2\I\pi}  
\det\left[\frac{1}{w_i+\la_i-w_j}\right]_{i,j=1}^{\ell(\lambda)}  \prod_{j=1}^{\ell(\lambda)} \frac{e^{t {\mathtt G}(w_j+\lambda_j) - \frac{x}{2} (w_j+\lambda_j)^2+ \frac{x}{2} (\lambda_j+w_j)}}{
e^{t {\mathtt G}(w_j) - \frac{x}{2} w_j^2+\frac{x}{2}w_j}}
\\
\times  \prod_{j=1}^{\ell(\lambda)} \frac{2^{-\lambda_j}  \Gamma \left(2(w_j+\lambda_j) -1 \right)}{ \Gamma (2 w_j+\lambda_j -1)} 
 \frac{\hat\eta(w_j+\lambda_j-1/2)}{\hat\eta(w_j-1/2)}  \\ \times  \prod_{1 \leq i < j \leq \ell(\lambda)} 
   \frac{\Gamma \left(w_i+w_j-1\right) \Gamma \left(w_i+w_j+\lambda _i+\lambda
   _j-1\right)}{\Gamma \left(w_i+w_j+\lambda _i-1\right) \Gamma \left(w_i+w_j+\lambda
   _j-1\right)}.
   \end{multline}
where
\begin{equation}
    \hat{\eta}(z) = \frac{\Gamma(v_1-z)}{\Gamma(v_1+z)}\frac{\Gamma(v_2-z)}{\Gamma(v_2+z)} \frac{1}{\Gamma(u+z)}.
\end{equation}
The contour for the variables $w_i$ has to be chosen so that 
\begin{equation}
\max\lbrace 1/2-u,1/2 \rbrace < a_w < v-1/2-\lambda_i+1
\label{eq:conditionscontours}
\end{equation}
for any $\lambda_i$ and $v=v_1, v_2$, and the moment formula was valid for $u+v>k$ and $v-\frac{1}{2}>k-1$, which implies that $u+v>\lambda_i, v-\frac{1}{2}> \lambda_i-1$ so that one can always find $a_w$ satisfying \eqref{eq:conditionscontours}. Summing to obtain the generating function, and using Mellin-Barnes the Mellin-Barnes integral representation, we obtain 
\begin{multline}
\EE\left[\exp \left(-\varsigma e^{\frac{t}{12}} W Z(x,t)\right) \right]\\
	= \sum_{\ell=0}^{+\infty} \frac{(-1)^{\ell}}{\ell!} 
  \int_{a_w+\I\R}\frac{\rmd w_1}{2\I\pi} 
 \dots \int_{a_w+\I\R}\frac{\rmd w_{\ell}}{2\I\pi}  
  \int_{\mathcal C_{a_s}[w_1]}\frac{\rmd s_1}{2\I\pi} 
 \dots \int_{\mathcal C_{a_s}[w_{\ell}]}\frac{\rmd s_{\ell}}{2\I\pi}  
\det\left[\frac{1}{s_i-w_j}\right]_{i,j=1}^{\ell}  \\
\times \prod_{j=1}^{\ell} \frac{e^{t {\mathtt G}(s_j) - \frac{x}{2} s_j^2+ \frac{x}{2} (s_j-w_j)}}{
e^{t {\mathtt G}(w_j) - \frac{x}{2} w_j^2}} \frac{\pi}{\sin(\pi(s_j-w_j))} (\varsigma e^{\frac{t}{12}})^{s_j-w_j} \frac{\hat \eta(s_j-1/2)}{\hat \eta(w_j-1/2)} \\
%&\times \prod_{j=1}^{\ell}  \frac{\Gamma (u+w_j)\Gamma (v_1+w_j)\Gamma (v_2+w_j) }{\Gamma (v_1-w_j+1)\Gamma (v_2-w_j+1) 
% }
% \frac{ \Gamma (v_1-s_j +1)\Gamma (v_2-s_j +1)}{\Gamma (u+s_j ) \Gamma (v_1+s_j )\Gamma (v_2+s_j )} 
 % \\
 \times  \prod_{j=1}^{\ell}
  \frac{  \Gamma \left(2 s_j -1 \right)}{ \Gamma (w_j+s_j -1)}  
  \prod_{1 \leq i < j \leq \ell} 
   \frac{\Gamma \left(w_i+w_j-1\right) \Gamma \left(s_i+s_j-1\right)}{\Gamma \left(s_i+w_j-1\right) \Gamma \left(w_i+s_j-1\right)}.
   \label{eq:MellinBarnesxpositif}
\end{multline}
The contour $\mathcal C_{a_s}[w]$ (depicted on Fig. \ref{fig:contourpluscircles}) is formed by two semi-infinite rays going to $\infty$ in the direction $\pm\pi/3$, starting from the horizontal axis at the point $a_s$ and the union of negatively oriented circles around the poles at $w+1, w+2, \dots$ when these lie to the left of the semi-infinite rays. The infinite part of the contour is oriented from bottom to top. 

The real numbers $a_s$ and $a_w$ has to be chosen so that (recall that after the analytic continuation, $\Re [\lambda] = \Re[s-w]=a_s-a_w$)
\begin{equation}
a_w<a_s<a_w+1, \quad\max\lbrace 1/2-u,1/2,1/2-v \rbrace < a_w, \quad a_s <v_1+1/2, \quad  u+v_1> a_s-a_w-1. 
\end{equation}
Note that the condition $a_s<a_w+1$ is not really needed since we have added the small circles to the contour for $s_i$. Additionally, we need to chose $a_s>1/2$, so that $\Gamma(2s-1)$ has no poles on the right of contours. One also needs to discuss the convergence of the $w$ integral. A sufficient condition for the integrals over $w_i$ to be convergent is that $a_w-1/2<x/2$. 

\subsection{Pfaffian formula in the large time limit}
We now consider the large time limit, and the critical region, hence we rescale 
\bea
&& u = a t^{-1/3}, \quad  \quad v_1 =  b t^{-1/3} , \quad \quad v_2 =  c t^{-1/3}, \\
&&  w_i \to  \frac{1}{2} + t^{-1/3} w_i, \quad \quad s_i \to  \frac{1}{2} + t^{-1/3} s_i,  \\
&& x = t^{2/3} \tilde x, \quad  \quad \varsigma = \exp(- y t^{1/3}). 
\eea 
Defining the rational function
\begin{equation}
    \eta(z)=(\Arescaled+z)\frac{(\Brescaled+z)(\Crescaled+z)}{(\Brescaled-z)(\Crescaled-z)}, 
\end{equation}
the right hand side becomes 
\begin{equation}
    \begin{split}
   &\EE\left[\exp \left(-\varsigma e^{\frac{t}{12}} W Z(x,t)\right) \right]\\ 
   &=\sum_{\ell=0}^{+\infty} \frac{(-1)^{\ell}}{\ell!} 
\int_{a_w+\I\R}\frac{\rmd w_1}{2\I\pi} 
\dots \int_{a_w+\I\R}\frac{\rmd w_{\ell}}{2\I\pi}  
\int_{\mathcal C_{a_s}}\frac{\rmd s_1}{2\I\pi} 
\dots \int_{\mathcal C_{a_s}}\frac{\rmd s_{\ell}}{2\I\pi}  
\det\left[\frac{1}{s_i-w_j}\right]_{i,j=1}^{\ell}  \\
&\times \prod_{j=1}^{\ell} \exp\left( \frac{s_j^3}{3}-\frac{w_j^3}{3} - \frac{\tilde x}{2}(s_j^2-w_j^2) -y (s_j-w_j) \right)
 \frac{s_j+w_j}{s_j-w_j} \frac{\eta(s_j)}{\eta(w_j)}\frac{1}{2s_j} \\ 
&\times \prod_{1 \leq i < j \leq \ell} \frac{(s_i+w_j)(s_j+w_i)}{(w_i+w_j)(s_i+s_j)}     
    \end{split}
\end{equation}

The contour $\mathcal C_{a_s}$ is now formed by two semi-infinite rays going to $\infty$ in the direction $\pm\pi/3$, starting from the horizontal axis at the point $a_s$ (without additional small circles). The values of $a_w, a_s$ now satisfy 
\begin{equation}
\max\lbrace -a,-b,-c, 0\rbrace  < a_w < a_s < \min\lbrace b,c\rbrace ,
\end{equation}
which can be happen only when $b,c>0$ and $a+b, a+c>0$. Note that the condition that $a_w, a_s$ are positive is important because of the denominators $(w_i+w_j)$ and $(s_i+s_j)$, and also the pole at $s=0$. Now, the condition for the convergence of the integrals over $w_i$ is that  $a_w < \frac{\tilde x}{2}$.

Observe that we may write 
    \begin{multline} 
        \prod_{j=1}^{\ell}(s_j+w_j)\prod_{1 \leq i < j \leq \ell}  \frac{(s_i+w_j)(s_j+w_i)}{(w_i+w_j)(s_i+s_j)}\det\left[\frac{1}{s_i-w_j}\right]_{i,j=1}^{\ell} \\
        =  \prod_{i=1}^{\ell} \frac{s_i+w_i}{s_i-w_i} \prod_{i<j} \frac{(s_i-s_j)(-w_i+w_j)(s_i+w_j)(-w_i-s_j)}{(-w_i-w_j)(s_i+s_j)(s_i-w_j)(-w_i+s_j)} \\
=\prod_{i<j} \frac{u_i-u_j}{u_i+u_j},
\end{multline}
where $\vec u = (s_1, -w_1, s_2, -w_2, \dots, s_\ell,-w_\ell)$.
We recognize Schur's Pfaffian formula, so that we obtain 
\begin{multline}
\EE\left[\exp \left(-\varsigma e^{\frac{t}{12}} W Z(x,t)\right) \right]= \\ \sum_{\ell=0}^{+\infty} \frac{(-1)^{\ell}}{\ell!} 
\int_{a_w+\I\R}\frac{\rmd w_1}{2\I\pi} 
\dots \int_{a_w+\I\R}\frac{\rmd w_{\ell}}{2\I\pi}  
\int_{\mathcal C_{a_s}}\frac{\rmd s_1}{2\I\pi} 
\dots \int_{\mathcal C_{a_s}}\frac{\rmd s_{\ell}}{2\I\pi}  
\Pf\left[\frac{u_i-u_j}{u_i+u_j}\right]_{i,j=1}^{2\ell}  \\
\times \prod_{j=1}^{\ell} \frac{\exp\left( \frac{s_j^3}{3}-\frac{w_j^3}{3} - \frac{\tilde x}{2}(s_j^2-w_j^2) -y (s_j-w_j) \right)}{(s_j-w_j)(2s_j)}
\frac{\eta(s_j)}{\eta(w_j)}
\end{multline}
This corresponds to 
\begin{align}
\EE\left[\exp \left(-\varsigma e^{\frac{t}{12}} W Z(x,t)\right) \right] &=\Pf[J-K]_{\mathbb L^2(0, +\infty)}\\
&=\sum_{\ell=0}^{+\infty} \frac{(-1)^{\ell}}{\ell!} \int_{\R_+}\rmd r_1 \dots \int_{\R_+} \rmd r_k \, \Pf\left[K(r_i,r_j)\right]_{i,j=1}^{\ell},
\label{eq:FredholmPfaffian}
\end{align}
where, after a change of variables $w \to - w$ in the integrals, we may define the kernel by 
\begin{subequations}
	\begin{align}
	K_{11}(r,r') &= \int_{\mathcal C_{a_s}} \frac{\rmd z}{2\I\pi} \int_{\mathcal C_{a_s}} \frac{\rmd  w}{2\I\pi}\frac{z-w}{z+w} e^{\frac{z^3}{3}+\frac{w^3}{3}-\frac{\tilde x}{2}(z^2+ w^2)-(r+y)z-(r'+y)w}\frac{\eta(z)\eta(w)}{4zw},\\
	K_{12}(r,r')&= \int_{\mathcal C_{a_s}} \frac{\rmd z}{2\I\pi} \int_{-a_w+\I\R} \frac{\rmd  w}{2\I\pi} \frac{z-w}{z+w} e^{\frac{z^3}{3}+\frac{w^3}{3}-\frac{\tilde x}{2}(z^2-w^2)-(r+y)z-(r'+y)w}\frac{\eta(z)}{2z\eta(-w)},\\
	K_{22}(r,r')&= \int_{-a_w+\I\R} \frac{\rmd z}{2\I\pi} \int_{-a_w+\I\R} \frac{\rmd  w}{2\I\pi} \frac{z-w}{z+w}e^{\frac{z^3}{3}+\frac{w^3}{3}-\frac{\tilde x}{2}(-z^2-w^2)-(r+y)z-(r'+y)w}\frac{1}{\eta(-z)\eta(-w)}.\label{eq:K22matrixkernel}
	\end{align}
\end{subequations}
We now shift the contour for $-a_w+\I\R$  to the right of $0$. In $K_{12}$ we can deform the contours without crossing any pole. In $K_{22}$, we do cross a pole and have to take into account the residue.
\begin{subequations}
	\begin{align}
	K_{11}(r,r') &= \int_{\mathcal C_{a_s}} \frac{\rmd z}{2\I\pi} \int_{\mathcal C_{a_s}} \frac{\rmd  w}{2\I\pi}\frac{z-w}{z+w} e^{\frac{z^3}{3}+\frac{w^3}{3}-\frac{\tilde x}{2}(z^2+ w^2)-(r+y)z-(r'+y)w}\frac{\eta(z)\eta(w)}{4zw},\\
	K_{12}(r,r')&= \int_{\mathcal C_{a_s}} \frac{\rmd z}{2\I\pi} \int_{\mathcal C_{a_s}} \frac{\rmd  w}{2\I\pi} \frac{z-w}{z+w} e^{\frac{z^3}{3}+\frac{w^3}{3}-\frac{\tilde x}{2}(z^2-w^2)-(r+y)z-(r'+y)w}\frac{\eta(z)}{2z\eta(-w)},\\
	K_{22}(r,r')&= \int_{\mathcal C_{a_s}} \frac{\rmd z}{2\I\pi} \int_{\mathcal C_{a_s}} \frac{\rmd  w}{2\I\pi} \frac{z-w}{z+w}e^{\frac{z^3}{3}+\frac{w^3}{3}-\frac{\tilde x}{2}(-z^2-w^2)-(r+y)z-(r'+y)w}\frac{1}{\eta(-z)\eta(-w)} \label{eq:K22matrixkernelbis}\\
	&-2\int_{\mathcal C_{a_s}} \frac{\rmd z}{2\I\pi}  e^{x z^2+(r'-r)z}\frac{z}{a^2-z^2}.
	\end{align}
\end{subequations}
The kernel has a particuliar structure. Define an operator $D$ such that for a function $f$ written as 
\begin{equation}
    f(r) = \int_{\mathcal C_{a_s}}  \frac{\mathrm d z}{2\I\pi}  \hat f(z)e^{-rz},
\end{equation}
then 
\begin{equation}
Df(r) =  \int_{\mathcal C_{a_s}} \frac{\mathrm d z}{2\I\pi} \hat D(z) \hat f(z)e^{-rz} 
\label{eq:actionD}
\end{equation}
where 
\begin{equation}
\hat D(z)=-2\frac{z e^{\tilde x z^2}}{a^2-z^2} =
    -e^{\tilde x z^2}\left[\frac{1}{a-z}-\frac{1}{a+z}\right].
\end{equation}
We have that 
\begin{equation}
    K= \begin{pmatrix}
    K_{11}&K_{12}\\K_{21}&K_{22} 
    \end{pmatrix} = \begin{pmatrix}
    K_{11} & -K_{11}D^{\intercal} \\ -D K_{11} & DK_{11}D^{\intercal} +D\varepsilon
    \end{pmatrix}
    \label{eq:Kstructured}
\end{equation}
where $\varepsilon$ is an operator with kernel 
$\varepsilon(r,r') = \delta(r-r').$ In other terms, $\varepsilon$ is the identity operator, i.e. $\varepsilon=I$.   The operator $D^{\intercal}$ acts on the left, i.e. $\bra{ f} D^{\intercal} \ket{ g } = \int_{0}^{\infty} \rmd  r Df(r) g(r)$. 

\subsection{From Fredholm Pfaffians to Fredholm determinants with scalar kernels}
\label{sec:appendixPfaffians}

Consider operators $B, D, \varepsilon: \mathbb{L}^2(\R_+) \to \mathbb{L}^2(\R_+) $. 
The operator  $D$ acts by multiplication in Fourier space (as in \eqref{eq:actionD}) and the operator  $\varepsilon$ is such that $D\varepsilon$ has an  anti-symmetric kernel. Recall the expression of the symplectic matrix 
\begin{equation}
J=
\begin{pmatrix}
0 & 1\\
-1 & 0
\end{pmatrix}.
\end{equation}
We can then perform the manipulations (similarly to \cite{krajenbrink2020painleve})
\begin{equation}
\begin{split}
{\rm Pf} \left(J-
\begin{bmatrix}
B&-BD^\intercal\\ 
-DB &  DBD^\intercal +D\varepsilon \end{bmatrix}\right)^2&={\rm Pf} \left(J-
\begin{bmatrix}
1 &0\\ 
0 &  D\end{bmatrix}
\begin{bmatrix}
B&-BD^\intercal\\ 
-B &  BD^\intercal +\varepsilon \end{bmatrix}\right)^2\\
&=\Det\left( I + \begin{bmatrix}
B &-BD^\intercal\\ 
-B & BD^\intercal+\varepsilon\end{bmatrix} J \begin{bmatrix}
1 &0\\ 
0 &  D\end{bmatrix}
\right)\\
&=\Det\left( I + \begin{bmatrix}
BD^\intercal &BD\\ 
-BD^\intercal -\varepsilon& -BD\end{bmatrix} 
\right).
\end{split}
\end{equation}
Summing the first line to the second one and subtracting the second column to the first one,  we obtain  

\begin{equation}
\begin{split}
\Det\left( I + \begin{bmatrix}
BD^\intercal &BD\\ 
-BD^\intercal-\varepsilon & -BD\end{bmatrix} 
\right)
&=\Det\left( I + \begin{bmatrix}
B(D^\intercal -D)&BD\\ 
-\varepsilon& 0\end{bmatrix} 
\right)\\
&=\Det\left(I+B(D^\intercal-D)+BD\varepsilon \right)\\
&=\Det\left(I+(D^\intercal-D)B+D\varepsilon B \right).
\end{split}
\end{equation}
To go from the first line with a matrix-valued kernel to the second line with a scalar kernel, we used a Schur's complement formula.
to go from the second line to the third one we used $\det(I+MN)=\det(I+NM)$.\\ 

For our specific kernel $K$ in  \eqref{eq:Kstructured}, we choose $B=K_{11}$ and $\varepsilon=I$, so that we obtain the simple formula
\begin{equation}
    \mathrm{Pf}(J-K) = \sqrt{ \det\left( I+ D^{\intercal} K_{11}\right)}=\sqrt{ \det\left( I+  K_{11}D^{\intercal}\right)},
\end{equation}
where now $K_{11}D^{\intercal}$ is a scalar kernel, acting on $\mathbb L^2(0,+\infty)$. 
Explicitly, we have 
\be \label{K11DT} 
(K_{11} D^\intercal)(r,r') = -  \int_{\mathcal C_{a_s}} \frac{\rmd z}{2\I\pi} \int_{\mathcal C_{a_s}} \frac{\rmd  w}{2\I\pi}\frac{z-w}{z+w} e^{\frac{z^3}{3}+\frac{w^3}{3}-\frac{\tilde x}{2}(z^2- w^2)-(r+y)z-(r'+y)w}\frac{\eta(z)}{2z\eta(-w)} 
\ee
with the rational factors
\begin{equation}
    \eta(z)=(\Arescaled+z)\frac{(\Brescaled+z)(\Crescaled+z)}{(\Brescaled-z)(\Crescaled-z)} 
\end{equation}

\subsection{Stationary limit}
As in \eqref{Feq} we want to calculate 
\be
G^{\rm HY}_{a,b}(y , \tilde x):= \lim_{c\to -b} \left(1 + \frac{\partial_y}{b+c}\right)  \sqrt{ \det\left( I+  K_{11}D^{\intercal}\right)}.
\ee 
in order to obtain the cumulative distribution of the height at large time for a stationary initial condition
with parameters $a,b$.

For now $a_s$ in \eqref{K11DT} satisfies $0<a_s<a,b,c$. To be able to perform the limit $c \to -b$ we move $c$ across the contours for $w$, $z$. It gives two additional residue terms
%\be 
%- {\rm Res}_{w=c} (K_{11} D^\intercal) =   \int_{\mathcal C_{a_s}} \frac{\rmd z}{2\I\pi}  e^{\frac{z^3}{3}+\frac{c^3}{3}-\frac{\tilde x}{2}(z^2- c^2)-(r+y)z-(r'+y)c}\frac{(a+z) (b+z)}{2z (b-z)} \frac{2 c (b+c)}{(a-c)(b-c)}
%\ee 
%\be 
%- {\rm Res}_{z=c} (K_{11} D^\intercal) =  - \int_{\mathcal C_{a_s}} \frac{\rmd w}{2\I\pi}  e^{\frac{w^3}{3}+\frac{c^3}{3}-\frac{\tilde x}{2}(c^2- w^2)-(r+y)c-(r'+y)w}
%\frac{b+w}{(a-w)(b-w)} \frac{(a+c)(b+c)}{b-c}
%\ee 
and no double residue. Define 
\begin{equation}
    f_{\pm}(r)=(a\pm c)^{\pm 1}\exp\left(\frac{c^3}{3}\mp \frac{\tilde x}{2} c^2-(r+y)c\right).
\end{equation}
\begin{equation} \label{defgp}
    g_+(r) = -\int_{\mathcal C_{a_s}} \frac{\rmd w}{2\I\pi} \exp(\frac{w^3}{3} +\frac{\tilde x}{2}w^2 -(r+y)w) \frac{(b+w)}{(b-w)(a-w)} 
\end{equation}
and 
\begin{equation} \label{defgm}
    g_-(r) = \int_{\mathcal C_{a_s}} \frac{\rmd z}{2\I\pi} \exp(\frac{z^3}{3} -\frac{\tilde x}{2}z^2 -(r+y)z)\frac{1}{z} \frac{(b+z)(a+z)}{(b-z)} 
\end{equation}
so that 
\begin{equation}
    - {\rm Res}_{z=c} (K_{11} D^\intercal) = \frac{b+c}{b-c} f_+(r)g_+(r'), \;\;\; - {\rm Res}_{w=c} (K_{11} D^\intercal) = c\frac{b+c}{b-c} f_-(r')g_-(r).
\end{equation}
Hence we may write 
\begin{equation}
    (K_{11} D^\intercal)(r,r') = L(r,r')+ \frac{b+c}{b-c}\left(   f_+(r)g_+(r') + c f_-(r')g_-(r)\right), 
\end{equation}
or in the operator formalism
\begin{equation}
    (K_{11} D^\intercal)= L + \frac{b+c}{b-c}\left(    \ket{f_+}   \bra{g_+}  + c  \ket{g_-}     \bra{f_-} \right), 
\end{equation}
where 
\begin{equation}
    L(r,r') = -  \int_{\mathcal C_{a_s}} \frac{\rmd z}{2\I\pi} \int_{\mathcal C_{a_s}} \frac{\rmd  w}{2\I\pi}\frac{z-w}{z+w} e^{\frac{z^3}{3}+\frac{w^3}{3}-\frac{\tilde x}{2}(z^2- w^2)-(r+y)z-(r'+y)w}\frac{\eta(z)}{2z\eta(-w)} 
\end{equation}
with both contours going between $\{0,c\}$ and $\{a,b\}$. Thus, we have 
\begin{equation}
    \det(I+K_{11} D^{\intercal}) = \det(I+L) \det\begin{pmatrix}
    1+ \frac{b+c}{b-c} \bra{ g_+ } \frac{1}{1+L} \ket{ f_+ } & \frac{b+c}{b-c}c\bra{ f_- } \frac{1}{1+L} \ket{  f_+ } \\ \frac{b+c}{b-c}\bra{ g_+ } \frac{1}{1+L} \ket{ g_- } & 1+ \frac{b+c}{b-c}  c \bra{f_- } \frac{1}{1+L} \ket{  g_- }.
    \end{pmatrix}
    \label{eq:limittotake}
\end{equation}
%We want to calculate
%\be
%\left(1 + \frac{\partial_y}{b+c}\right)  \sqrt{ \det\left( I+ D^{\intercal} K_{11}\right)}
%\ee 
Letting $\epsilon=b+c$, we will write upon expanding in $\epsilon$
\be 
\eqref{eq:limittotake} \simeq ( \det (I + L_0 + \epsilon X ) (C_0 + C_1 \epsilon + C_2 \epsilon^2) 
\ee 
and show that  $C_0= C_1= 0$, so that 
\be 
G^{\rm HY}_{a,b}(y , \tilde x) = \lim_{\epsilon \to 0}   \left(1 + \frac{\partial_y}{b+c}\right) \sqrt{\det(I+K_{11} D^{\intercal}) } = \partial_y \sqrt{ \det (I + L_0) C_2} \, .
\ee 
Let us calculate some inner products explicitly and expand in $\epsilon$ each term in \eqref{eq:limittotake}
\begin{itemize}
    \item One has
\be 
\frac{b+c}{b-c}c\braket{ f_- \vert  f_+ } = \frac{1}{2} \frac{b+c}{b-c} \frac{a+c}{a-c}  e^{\frac{2}{3} c^3 - 2 y c} 
\ee 
This scalar product is defined only for $c>0$ but we will consider the analytic continuation to $-b <c <0$. 
\item One has
\begin{equation}
\begin{split}
\frac{b+c}{b-c}  c \braket{f_- \vert   g_- }
&= \frac{b+c}{b-c} \frac{c}{a-c} e^{\frac{c^3}{3} + \frac{\tilde x}{2} c^2 - y c} 
\int_{\mathcal C_{a_s}} \frac{\rmd z}{2\I\pi} e^{\frac{z^3}{3} -\frac{\tilde x}{2}z^2 - y z}\frac{1}{z} \frac{(b+z)(a+z)}{(b-z)(z+c)} \\
&= -1 + \epsilon \left(b \tilde x + y + \frac{1}{2 b} - b^2 - \frac{1}{a+b}\right)  \\
& +  \frac{\epsilon}{2} \frac{1}{a+b} e^{- \frac{b^3}{3} + \frac{\tilde x}{2} b^2 + y b} 
\int \frac{\rmd z}{2\I\pi} e^{\frac{z^3}{3} -\frac{\tilde x}{2}z^2 - y z}\frac{1}{z} \frac{(b+z)(a+z)}{(z-b)^2}
\end{split}
\end{equation}
where we have moved the contour to $z>b$ and taken the residue, which for $c=-b$ has a pole. The remaining integral has no singularity at $c=-b$. 
\item One has
\begin{equation}
\begin{split}
\frac{b+c}{b-c} \braket{ g_+ \vert f_+ } &= - \frac{b+c}{b-c}  
(a+ c) e^{ \frac{c^3}{3} -  \frac{\tilde x}{2} c^2- y c}
\int \frac{\rmd w}{2\I\pi} e^{\frac{w^3}{3} +\frac{\tilde x}{2}w^2 - y w} \frac{(b+w)}{(b-w)(a-w)(w+c)} \\
&= -1+ \epsilon \left(y  - \frac{1}{2b} +\frac{1}{b-a}-b(b+\tilde x)\right) \\ 
&+ \epsilon \frac{a-b}{2b}e^{-\frac{b^3}{3}-\frac{\tilde x}{2}b^2+yb} \int \frac{\rmd w}{2\I\pi} e^{\frac{w^3}{3}+\frac{\tilde x}{2}w^2-yw}\frac{b+w}{(a-w)(w-b)^2}
\end{split}
\end{equation}
\item One has 
\begin{equation}
\begin{split}
& \frac{b+c}{b-c}\braket{ g_+  \vert g_- } \\
& = - \frac{b+c}{b-c}
\int_{\mathcal C_{a_s}} \frac{\rmd w}{2\I\pi}
\int_{\mathcal C_{a_s}} \frac{\rmd z}{2\I\pi} \frac{1}{w+z} 
e^{ \frac{w^3}{3} +\frac{\tilde x}{2}w^2 - y w + \frac{z^3}{3} -\frac{\tilde x}{2}z^2 - y z} 
\frac{(b+w)(b+z)(a+z)}{(b-w)(a-w)(b-z)}  \frac{1}{z}  
\end{split}
\end{equation} 
\end{itemize}

Let 
\begin{equation} \label{defL0} 
\begin{split}
    L^0(r,r') &=\lim_{\epsilon\to 0} L(r,r') \\
    &= -  \int \frac{\rmd z}{2\I\pi} \int \frac{\rmd  w}{2\I\pi}\frac{z-w}{z+w} e^{\frac{z^3}{3}+\frac{w^3}{3}-\frac{\tilde x}{2}(z^2- w^2)-(r+y)z-(r'+y)w}\frac{a+z}{2z(a-w)},
    \end{split}
\end{equation}
where the contours pass between $0$ and $a$.   We can see that $L^0$ can be bounded, for any $M>0$, as $\vert L_0(r,r') \vert \leqslant C e^{-Mr-ar'}$
for some constant $C$. Let us also define 
\begin{equation} \label{deff0}
      f_{\pm}^0(r) = \lim_{\epsilon\to 0}  f_{\pm}(r)=(a\mp b)^{\pm 1}\exp\left(-\frac{b^3}{3}\mp \frac{\tilde x}{2} b^2+(r+y)b\right).
\end{equation}
We compute now the inner products in \eqref{eq:limittotake}. Since the functions $g^+$ and $g^-$ decay exponentially fast at infinity, we simply have 
\begin{equation}
    \frac{b+c}{b-c}\bra{ g_+ } \frac{1}{1+L} \ket{ g_- } = \frac{\epsilon}{2b} \bra{ g_+ } \frac{1}{1+L_0} \ket{ g_-}+ o(\epsilon).
\end{equation}
To compute the remaining inner products, we use the  decomposition 
$$ \frac{1}{1+L} = 1 -\frac{L}{1+L}.$$ 
\begin{itemize}
    \item One has
    \be 
\frac{b+c}{b-c}c \bra{ f_- } \frac{1}{1+L} \ket{  f_+ } =   \frac{\epsilon}{4b} \frac{a-b}{a+b}  e^{-\frac{2}{3} b^3 + 2 y b} + \frac{\epsilon}{2} \bra{ f_-^0} \frac{L^0}{1+L^0}  \ket{  f_+^0 } + o(\epsilon).
\ee 
\item One has
\begin{multline} 
    \frac{b+c}{b-c} c\bra{ f_- } \frac{1}{1+L} \ket{   g_- }  = -1 + \epsilon \left(b \tilde x + y + \frac{1}{2 b} - b^2 - \frac{1}{a+b}\right)  \\
+  \frac{\epsilon}{2} \frac{1}{a+b} e^{- \frac{b^3}{3} + \frac{\tilde x}{2} b^2 + y b}
\int \frac{\rmd z}{2\I\pi} e^{\frac{z^3}{3} -\frac{\tilde x}{2}z^2 - y z} \frac{(b+z)(a+z)}{z(z-b)^2} +\frac{\epsilon}{2} \bra{  f_-^0 } \frac{L^0}{1+L^0} \ket{   g_- } +o(\epsilon)
\end{multline}
where the contour passes to the right of $b$.
\item One has 
\begin{multline}
    \frac{b+c}{b-c} \bra{ g_+  } \frac{1}{1+L}\bra{ f_+ } =  -1+ \epsilon \left(y  - \frac{1}{2b} +\frac{1}{b-a}-b(b+\tilde x)\right) \\ 
+ \epsilon \frac{a-b}{2b}e^{-\frac{b^3}{3}-\frac{\tilde x}{2}b^2+yb} \int \frac{\rmd w}{2\I\pi} e^{\frac{w^3}{3}+\frac{\tilde x}{2}w^2-yw}\frac{b+w}{(a-w)(w-b)^2} -\frac{\epsilon}{2b} \bra{ g_+  } \frac{L^0}{1+L^0}\ket{ f_+^0 }+ o(\epsilon),
\end{multline}
where the contour passes to the right of $b$ and to the left of $a$. 
\end{itemize}

Hence our final result is that for the initial condition 
$h(x,0) = \mathcal{HY}_{u,v}(x)$, where  $u=at^{-1/3}, v=bt^{-1/3}$, with $a+b>0, b\leq 0$
\begin{equation}
    \lim_{t\to \infty} \mathbb P\left( \frac{h_{at^{-1/3}}(t^{2/3}\tilde x,t)+\frac{t}{12} }{ t^{1/3} } \leqslant y\right)  = G_{a,b}^{\rm HY}(y, \tilde x).
\end{equation}
The CDF of the solution with Hariya-Yor initial condition is
\be 
G^{\rm HY}_{a,b}(y , \tilde x) =  \partial_y \sqrt{ \det (I + L_0) \det(M)} 
\label{eq:defGforxpositif}
\ee
where $M$ is the $2$ by $2$ matrix 
\begin{equation}
M=\begin{pmatrix}
M_{11} & M_{12} \\ M_{21}& M_{22}
\end{pmatrix},
\end{equation}
with 
\begin{align}
M_{11}&=  y  - \frac{1}{2b} +\frac{1}{b-a}-b(b+\tilde x)  -\frac{1}{2b} \bra{ g_+  } \frac{L^0}{1+L^0}\ket{ f_+^0 } \\ 
& \hspace{2cm}  +\frac{a-b}{2b}e^{-\frac{b^3}{3}-\frac{\tilde x}{2}b^2+yb} \int \frac{\rmd w}{2\I\pi} e^{\frac{w^3}{3}+\frac{\tilde x}{2}w^2-yw}\frac{b+w}{(a-w)(w-b)^2}, \nonumber \\ 
M_{12}&=  \frac{1}{4b} \frac{a-b}{a+b}  e^{-\frac{2}{3} b^3 + 2 y b} + \frac{1}{2} \bra{ f_-^0 } \frac{L^0}{1+L^0}  \ket{  f_+^0}, \\
M_{21}&= \frac{1}{2b} \bra{ g_+ } \frac{1}{1+L_0} \ket{ g_-},  \\
M_{22}&=   b \tilde x + y + \frac{1}{2 b} - b^2 - \frac{1}{a+b} +\frac{1}{2} \bra{  f_-^0 } \frac{L^0}{1+L^0} \ket{   g_- }  \\ &\hspace{3cm}   +  \frac{e^{- \frac{b^3}{3} + \frac{\tilde x}{2} b^2 + y b}}{2(a+b)} 
\int \frac{\rmd z}{2\I\pi} e^{\frac{z^3}{3} -\frac{\tilde x}{2}z^2 - y z}\frac{(b+z)(a+z)}{z(z-b)^2}. \nonumber 
\end{align}
%\begin{multline} 
%    M = \\ 
%   \left[\begin{array}{c|c}
%   \begin{matrix} y  - \frac{1}{2b} +\frac{1}{b-a}-b(b+\tilde x) \\ +\frac{a-b}{2b}e^{-\frac{b^3}{3}-\frac{\tilde x}{2}b^2+yb} \int \frac{\rmd w}{2\I\pi} e^{\frac{w^3}{3}+\frac{\tilde x}{2}w^2-yw}\frac{b+w}{(a-w)(w-b)^2} \\ -\frac{1}{2b} \bra{ g_+  } \frac{L^0}{1+L^0}\ket{ f_+^0 }\end{matrix} &  
%     \frac{1}{4b} \frac{a-b}{a+b}  e^{-\frac{2}{3} b^3 + 2 y b} + \frac{1}{2} \bra{ f_-^0 } \frac{L^0}{1+L^0}  \ket{  f_+^0}
%    \\
%    \hline 
%     \frac{1}{2b} \bra{ g_+ } \frac{1}{1+L_0} \ket{ g_-}
%    & 
%  \begin{matrix}   b \tilde x + y + \frac{1}{2 b} - b^2 - \frac{1}{a+b}  \\ +  \frac{e^{- \frac{b^3}{3} + \frac{\tilde x}{2} b^2 + y b}}{2(a+b)} 
%\int \frac{\rmd z}{2\I\pi} e^{\frac{z^3}{3} -\frac{\tilde x}{2}z^2 - y z}\frac{(b+z)(a+z)}{z(z-b)^2}\\ +\frac{1}{2} \bra{  f_-^0 } \frac{L^0}{1+L^0} \ket{   g_- } \end{matrix} 
%\end{array}
%\right]
%\end{multline}
We recall that $L_0$ is defined in \eqref{defL0}, $f_\pm^0$ in \eqref{deff0}, $g^\pm$ in \eqref{defgp}, \eqref{defgm}.
Although this result is fully explicit, it is quite involved and it remains to be studied how the various known limits can
be obtained from it.

\bibliographystyle{unsrt} 
\bibliography{biblioHY.bib}

\end{document}